\documentclass[preprint2]{aastex}

\begin{document}
\title{Solitary Magnetic Bubbles} 
\author{Andrei Gruzinov}
\affil{CCPP, Physics Department, New York University, 4 Washington Place, New York, NY 10003}

\begin{abstract}

Stability and attractor property of free-floating axisymmetric magnetic bubbles in  high-conductivity plasmas is (tentatively, numerically) demonstrated. The existence of compact non-singular axisymmetric magnetic equilibria is proved. Being attractors, the solitary magnetic bubbles should exist in nature.

~
~

\end{abstract}

\section{Introduction}

We show that compact non-singular axisymmetric equilibrium magnetic field configurations are attractors (more precisely intermediate asymptotics) in a  weakly dissipative plasma. This means that some initial magnetic field configurations evolve into solitary magnetic bubbles. 

Because of the attractor property, such free-floating solitary magnetic bubbles should exist in nature (see Braithwaite 2010 and references therein).

\begin{figure}
\plottwo{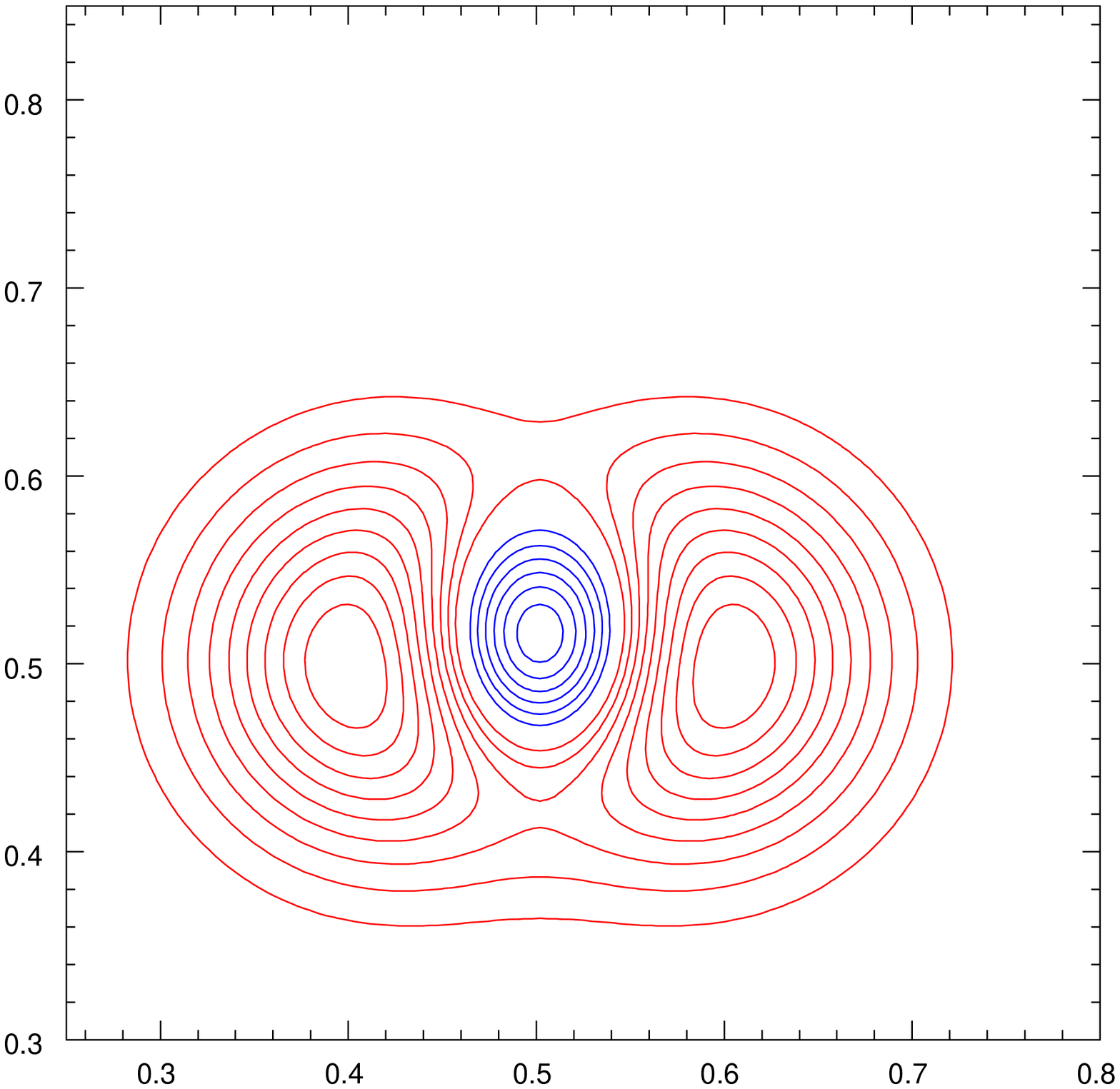}{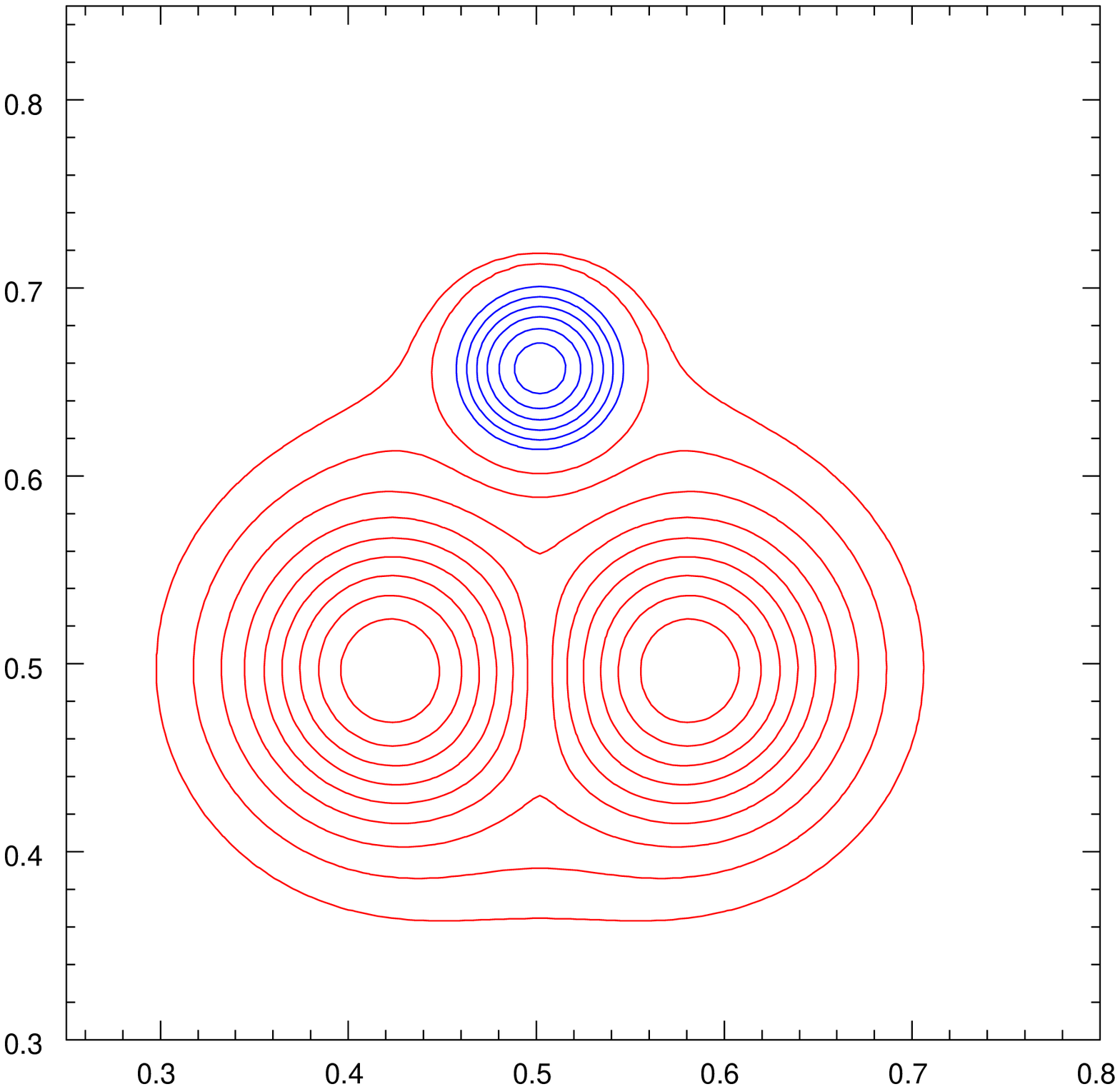}
\plottwo{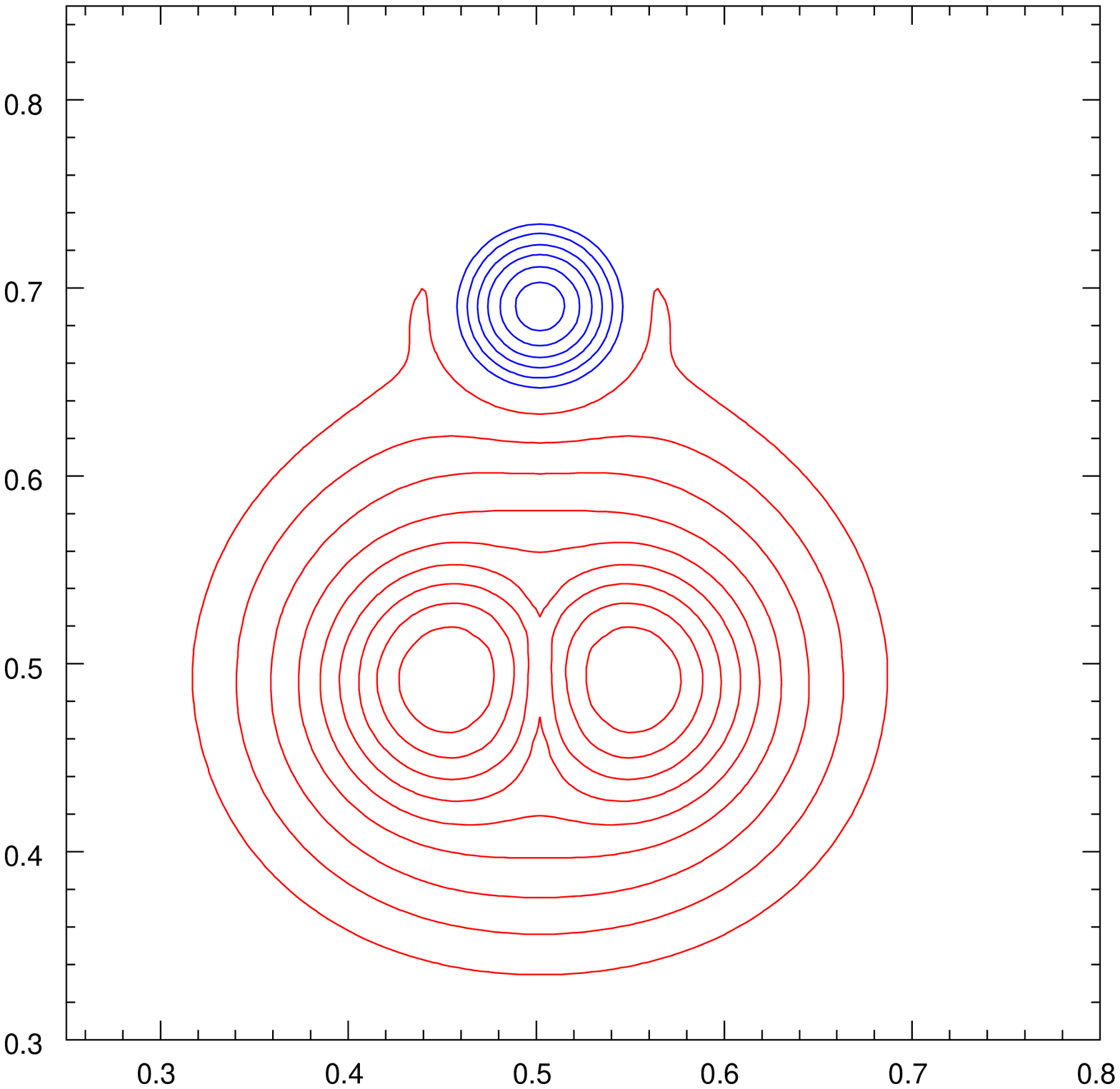}{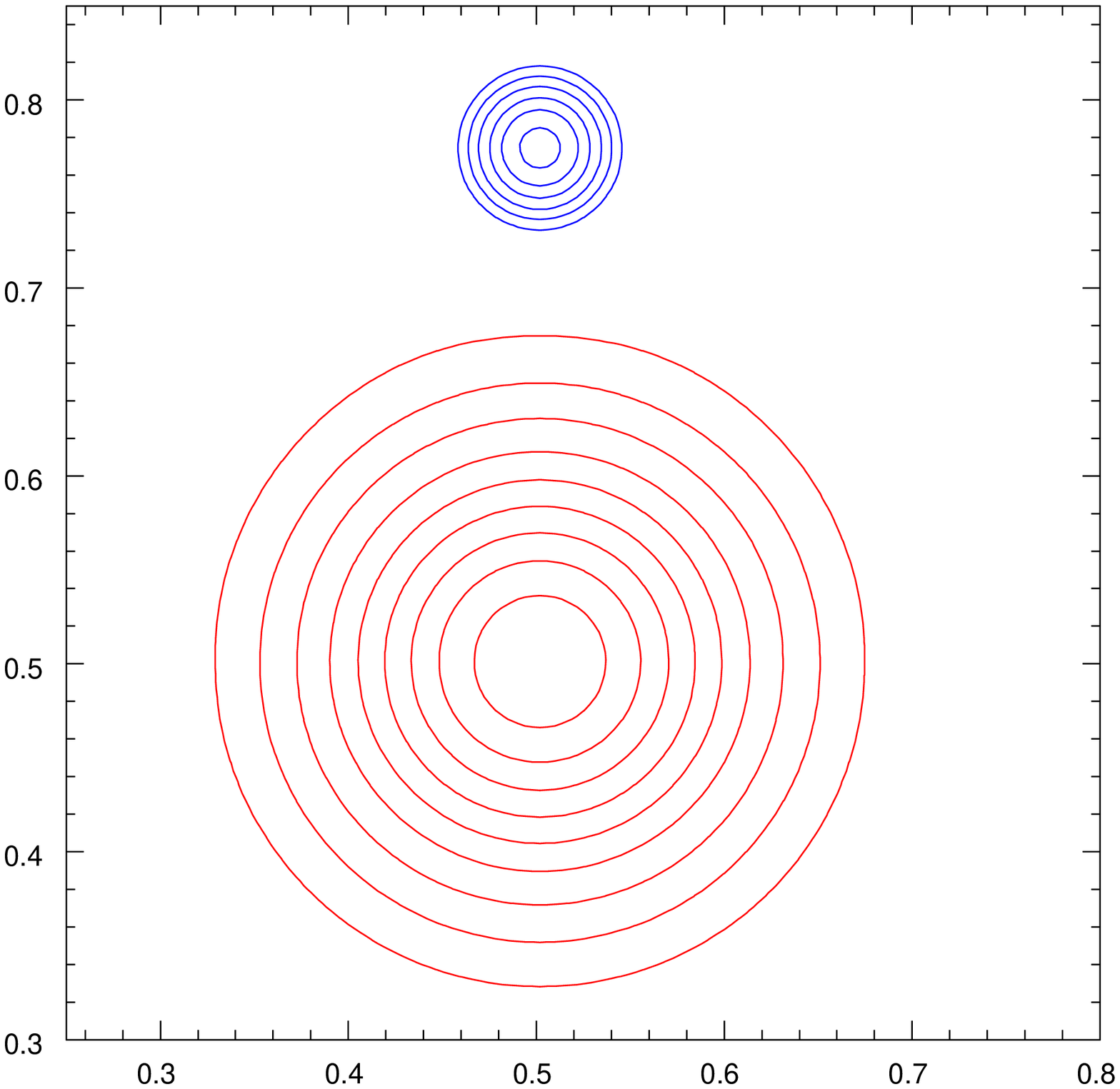}

\caption{{\it Upper Left}: Initial isolines of $\psi$. Different colors indicate different signs of $\psi$. {\it Upper Right}: Singular equilibrium stage. {\it Lower Left}: Reconnection stage. {\it Lower Right}: Non-singular equilibrium stage }
\end{figure}

\section{Generic and intermediate-asymptotic magnetic equilibria}

First consider magnetic equilibria of ideally conducting incompressible plasmas. The magnetic field ${\bf B}$ will be in equilibrium if the magnetic ${\bf j}\times {\bf B}$ force is balanced by the pressure gradient:
\begin{equation}\label{equi}
{\bf B}\times \nabla \times {\bf B}=-\nabla p. 
\end{equation}
It turns out, that due to topological obstacles (Arnold 1986), generic magnetic equilibria, that is generic solutions of eq.(\ref{equi}), contain singular current layers (Gruzinov 2009\footnote{The main prediction of this paper -- generic equilibrium has a dense set of singular current layers -- has been confirmed (tentatively, numerically)}). 

If the plasma electric conductivity is finite, but still very high, the magnetic diffusivity will disperse the singular current layers by the reconnection process, without dispersing the entire magnetic field configuration. This is the intermediate-asymptotic stage we are interested in. During the intermediate-asymptotic stage, the magnetic equilibrium configuration should be non-singular. 

Ideally, we want to find all compact non-singular magnetic equilibria. We think we have a full solution of this problem in two dimensions (\S3 ). In 2D, the only compact non-singular equilibrium is a circular-symmetric magnetic bubble.

In 3D, there is a class of non-singular equilibrium configurations -- compact axisymmetric magnetic bubbles (\S4 ). The non-singular axisymmetric equilibrium bubbles are attractors (\S5 ). But unlike the 2D case, we do not know if these bubbles are the only non-singular compact attractors.

We study magnetic equilibria using the relaxation method (e.g. Gruzinov 2009). We put an arbitrary compact magnetic field into high-viscosity small-magnetic-diffusivity plasma, and let the plasma move under the resulting magnetic stress:

\begin{equation}\label{relax}
\begin{array}{c}
\dot{{\bf B}}=\nabla \times ({\bf v}\times {\bf B})+\eta \Delta {\bf B},\\ 
{\bf v}=-\Delta ^{-2}\nabla \times \nabla \times ({\bf B}\times \nabla \times {\bf B}),
\end{array}
\end{equation}

The first equation (with $\eta =0$) says that the magnetic field ${\bf B}$ is frozen into the fluid moving with velocity ${\bf v}$. Non-zero $\eta$ describes magnetic diffusivity, $\Delta \equiv \nabla ^2$. The second equation says that the fluid incompressibly yields to the magnetic force. The equations are solved numerically, using the fast Fourier transformation.

Needless to say, the creeping flow evolution is not what happens in nature. But the final non-singular equilibria that we find should be real. The idea is that the route to the equilibrium is irrelevant, we just find generic non-singular equilibria.

\section{Generic and intermediate-asymptotic magnetic equilibria in 2D}

Consider 2D magnetic fields, that is magnetic fields of the form

\begin{equation}\label{2Dfield}
{\bf B}=(-\partial _y\psi, \partial _x \psi ,0), ~~~\psi=\psi(t,x,y).
\end{equation}

Take a generic initial magnetic stream function $\psi$ vanishing at spatial infinity and let it evolve according to eq.(\ref{relax}) with zero magnetic diffusivity, $\eta =0$. Then each saddle of $\psi$ develops into a singular current layer (Gruzinov 2009).

If we want to have a non-singular equilibrium, we must start with the initial $\psi$ without saddles. Then the isolines of $\psi$ are nested closed curves. The ideal relaxation of magnetic energy simply turns all the nested curves into concentric circles, giving the final state $\psi =\psi (r)$ where $r$ is the distance from some center. 

Now consider slightly non-ideal relaxation, with magnetic diffusivity $\eta >0$ but still very small. Numerical experiments (some 300 different runs) show that the generic initial state with compact support evolves as follows:

\begin{itemize}

\item Sharp current layers form at saddle points

\item Reconnection occurs at current layers

\item The field splits into a number of circular magnetic bubbles.

\item The bubbles have monotonic $\psi (r)$, with a maximum or minimum at $r=0$ and vanishing  $\psi$ at large $r$. 

\item Bubbles of the same polarity (same sign of $\psi$) merge, bubbles of the opposite polarity repel each other.

\item Ultimately, two bubbles form, one positive and one negative, and the distance between the two final bubbles is gradually increasing.

\end{itemize}

This is illustrated in fig.1. At high numerical resolution, when the small magnetic diffusivity can be achieved, the different stages of the relaxation process are easily distinguished. At first, ideal relaxation quickly moves the field into the minimum-energy-for-a-given-topology state. Then reconnection occurs, on the time scale which is much longer than the ideal relaxation time. After that, an almost steady non-singular equilibrium state is established.

The 2D summary is: (i) generic equilibrium is singular, (ii) the only non-singular compact equilibrium configurations are circular bubbles, (iii) for arbitrary compact initial field, circular solitary magnetic bubbles are the only attractors (more precisely, intermediate asymptotics of the weakly dissipative MHD).

This behavior -- formation of two infinitely separated circular bubbles -- seems too simple to be true. But in fact its origin is quite clear. In 2D, there are infinitely many robust invariants, which are constant even when the magnetic energy can decrease due to magnetic diffusivity. From eq.(\ref{relax}),
\begin{equation}\label{2Dinv}
{d\over dt}\int d^2r~F(\psi )=-\eta \int d^2r~F''(\psi )(\nabla \psi )^2.
\end{equation}
This is very small, because the rate of dissipative magnetic energy damping is given by the integral of $\eta (\Delta \psi)^2$. 

The final state is simply the state with minimal energy for the given area per each interval $d\psi$. This minimum is given by the state with two monotonic circular bubbles, one positive and one negative.

\begin{figure}
\plottwo{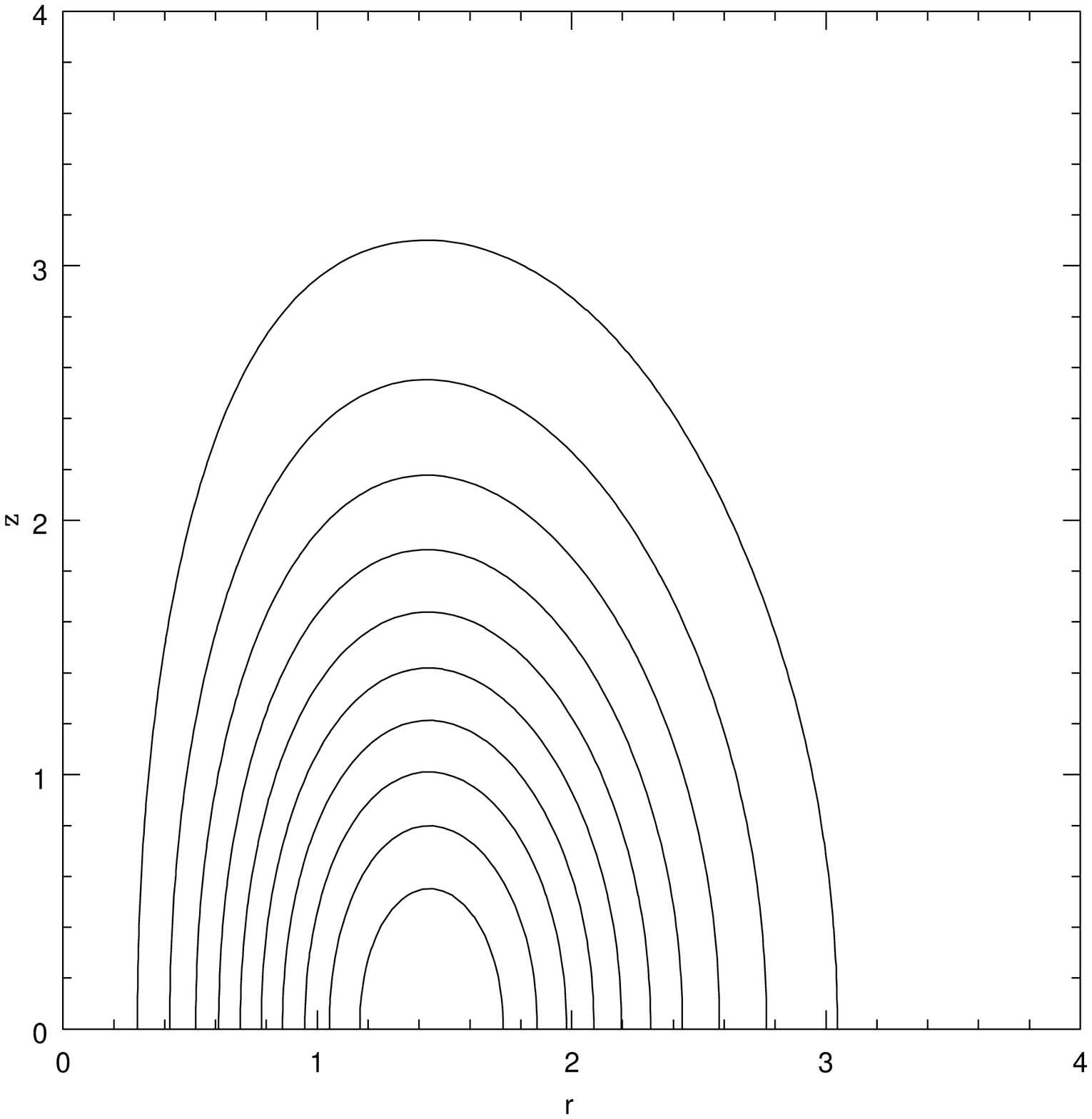}{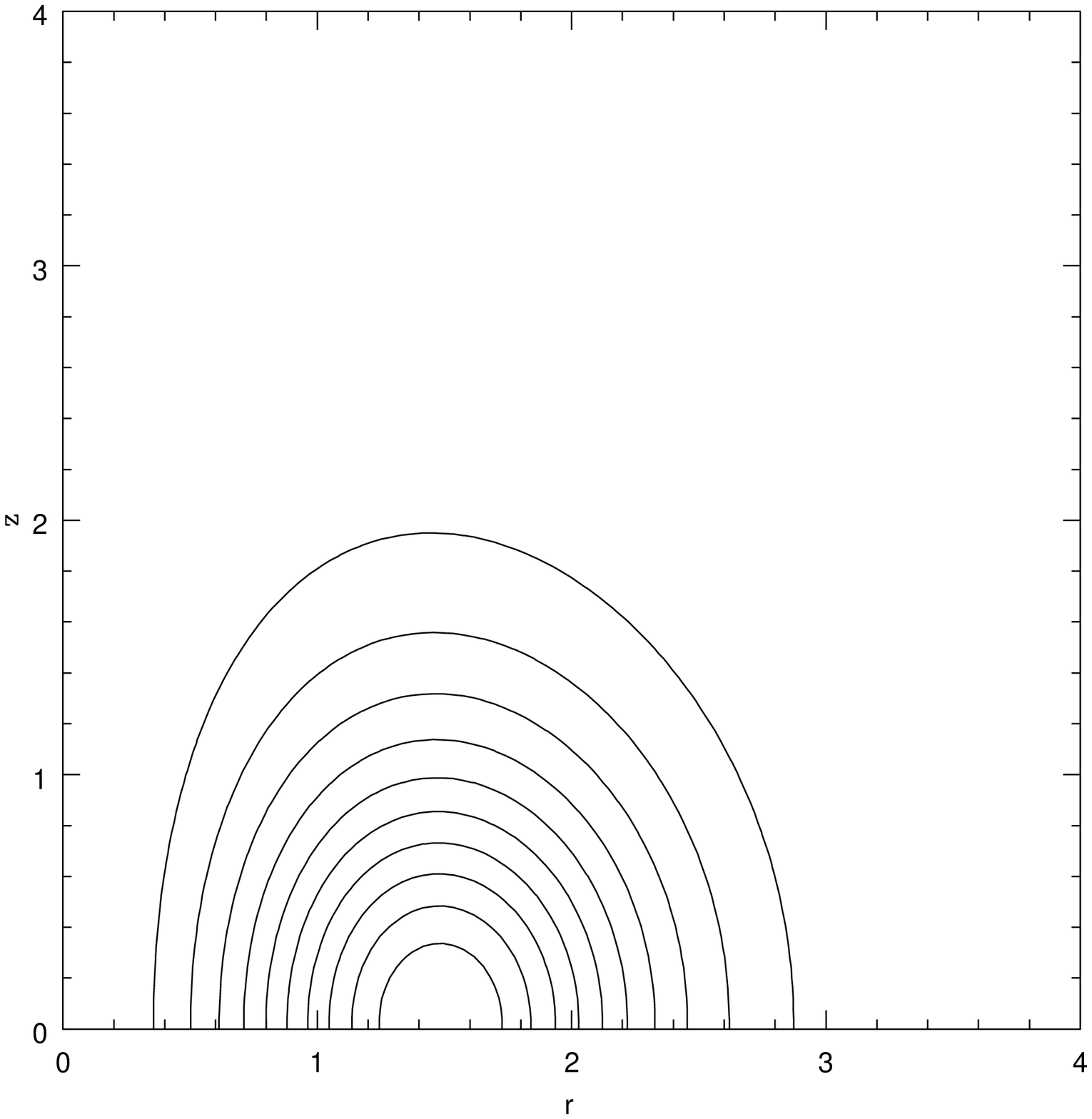}

\caption{ {\it Left}: Isolines of $\psi$, from eq.(\ref{GSex}) with $\lambda =2.74$. {\it Right}: $\lambda=0.523$ }
\end{figure}

\section{Axisymmetric non-singular magnetic equilibria}

Now we want to find the weakly-dissipative 3D attractors. Unlike the 2D case, we cannot claim a full understanding of this problem. This is because the runs take longer, and the magnetic fields are harder to visualize in 3D. Also, in 2D, we know that all saddles become singular current layers under the ideal relaxation. In 3D, generic ideal equilibria are also singular, with infinitely many singular current layers (Gruzinov 2009), but it remains unclear where the current layers form.

However one thing seems to emerge from our numerical experiments with the weakly dissipative magnetic relaxation in 3D (about 100 different runs). There exist non-singular axisymmetric compact magnetic configurations, which are attractors. In this section we describe the non-singular axisymmetric equilibrium bubbles, and in \S5 we demonstrate their attractor and stability properties.

In cylindrical coordinates $(r, \theta, z)$, the axisymmetric magnetic field can be written as follows
\begin{equation}\label{3Dfield}
{\bf B}={1\over r}(-\partial _z\psi, A, \partial _r\psi ),
\end{equation}
where $\psi=\psi (r,z)$, $A=A(r,z)$. Then the current is
\begin{equation}\label{curl}
\nabla \times {\bf B}={1\over r}(-\partial _zA, -\Delta _*\psi, \partial _rA ),
\end{equation}
where $\Delta _* \equiv \partial _r^2-{1\over r}\partial _r+\partial _z^2$. The equilibrium equation (\ref{equi}) then reads
\begin{equation}\label{aequi1}
\Delta _* \psi \partial _r\psi+A\partial _rA+r^2\partial _rp=0,
\end{equation}
\begin{equation}\label{aequi2}
\Delta _* \psi \partial _z\psi+A\partial _zA+r^2\partial _zp=0,
\end{equation}
\begin{equation}\label{aequi3}
\partial _r\psi\partial _zA-\partial _z\psi\partial _rA=0.
\end{equation}

From eq.(\ref{aequi3}), $A=A(\psi )$. Then from eqs.(\ref{aequi1} ,\ref{aequi2}), $p=p(\psi )$, and we get the Grad-Shafranov equation
\begin{equation}\label{GS}
\Delta _*\psi+AA'+r^2p'=0.
\end{equation}
where $'\equiv {d\over d\psi }$.

This equation describes axisymmetric magnetic configurations which minimize magnetic energy
\begin{equation}\label{energy}
E=\int dr~dz~{1\over r}((\nabla \psi )^2+A^2)
\end{equation}
for a fixed volume between magnetic surfaces 
\begin{equation}\label{volume}
I[f] =\int dr~dz~rf(\psi )={\rm const}, ~~~\forall f,
\end{equation}
and a fixed toroidal flux between magnetic surfaces
\begin{equation}\label{flux}
J[g]=\int dr~dz~{1\over r}g(\psi )A={\rm const}, ~~~\forall g.
\end{equation}
This variational formulation of the Grad-Shafranov equation suggests that non-singular solutions should exist for the trivial topology of magnetic surfaces. The trivial topology here means the absence of separatrix surfaces. The magnetic surfaces should be nested tori with a single maximum line. 

The invariant functionals $I$ and $J$, being linear, can be replaced, without loss of generality, by some invariant functions of a single variable, say $I(a)\equiv I[\delta (\psi -a)]$, $J(a)\equiv J[\delta (\psi -a)]$ -- the volume and the toroidal flux distributions functions. Thus, we see that the axisymmetric magnetic equilibria are parametrized by two functions of a single variable.

In the 2D case, we only have one invariant function $I$. The final state is also a function of a single variable, $\psi (r)$. Thus, it is not surprising that the 2D non-singular equilibria are all circular symmetric. In the 3D axisymmetric case, we have two constraints, and therefore the shape of the non-singular solution cannot be universal. If one keeps $I$ fixed, and changes $J$, the tori of constant $\psi$ will incompressibly deform.

The Grad-Shafranov equation (\ref{GS}) can be easily solved numerically (and even analyzed theoretically for some $A$ and $p$). Perhaps the simplest non-trivial case (coming from $p=-{1\over 2}\psi ^2$, ${1\over 2} A^2={1\over 2} \lambda\psi ^2+ {1\over 3} \psi ^3$) is 
\begin{equation}\label{GSex}
-\Delta _*\psi+V\psi =\lambda \psi
\end{equation}
Here $V=r^2-\psi$ is the potential of the Shrodinger equation, and $\lambda$ is the energy level. The potential is confining in the $r$ direction. Confinement along $z$ is self-generated by positive $\psi$. It is clear that non-singular solutions exist, at least for $\lambda$ slightly smaller than the ground state energy of the radial part of the hamiltonian ( the operator -$\partial _r^2+{1\over r}\partial _r+r^2$).

To confirm the non-singularity, and for the stability studies of \S5, we have solved eq.(\ref{GSex}) numerically, using the relaxation method 
\begin{equation}\label{GSexn}
\psi \rightarrow C\cdot (\psi +dt(\Delta _*\psi+\psi ^2-r^2\psi)),
\end{equation}
where the time step $dt$ should be small enough, and the normalization factor $C$ at each time step is chosen so as to keep the maximum value of $\psi$ at a pre-fixed value. The value of $\lambda$ is given by the asymptotic value of $C$, $C=1+\lambda dt$. The results are shown in fig.(2).

In summary, a family of non-singular axisymmetric equilibria parametrized by two functions of a single variable exists.

\section{Stability and the attractor property.}

Are the axisymmetric non-singular bubbles stable in 3D? Are these bubbles attractors of the weakly dissipative MHD?

\subsection{Stability}

Stability of plasma in magnetic fields is the most elaborate part of plasma physics (Kadomtsev 1966). The development of the MHD stability theory was motivated by the practical problem of confining the nuclear burning plasma by external magnetic fields.

In astrophysics, the MHD stability theory has found another application. Instead of asking which external magnetic configurations will stably confine the plasma, one asks what magnetic configurations can be stably supported by the plasma. What are the possible relic magnetic fields in non-convective stars, or what kinds of magnetic relics can exist in the ISM?

This problem (formulated by Chandrasekhar and Fermi 1953) is still unsolved. One can show that certain magnetic field configurations are unstable (Tayler 1982 and references therein). But the proof of stability is more difficult. One needs to show that the energy functional is non-decreasing under an {\it arbitrary} virtual displacement, and this turned out to be hard (although Gruzinov 2008 proves stability for some artificial magnetic configuration).

A possible way to check the stability of a given magnetic field numerically is by the ideal (meaning $\eta=0$) relaxation method (\ref{relax}). If the equilibrium is the energy saddle or maximum, rather than minimum, the relaxation will destroy it. We checked that the relaxation scheme does catch the classical kink and screw instabilities (for definitions, see Kadomtsev 1966)

When we take the equilibrium bubble (\ref{GSex}) with $\lambda =2.74$ (shown in fig.2), add a random weak magnetic field (to seed the possible instabilities), and subject it to the 3D relaxation (\ref{relax}), we see just the relaxation of the random component (magnetic energy decreases a bit). We also see that the noise slightly tilts the bubble (fig.3).

Of course we cannot conclude that the bubble is stable, especially given the danger of resonant surfaces where the field lines are closed. But the instabilities, if any, should be mild and globally non-destructive.

\subsection{Attractor property}

Given that the non-singular axisymmetric bubbles are stable, it is tempting to suggest that they behave just like the circular 2D bubbles. This is true, except for one major difference. Recall that the opposite polarity 2D bubbles repel, and the same polarity bubbles merge. In 3D, the role of polarity is played by the magnetic helicity $H\equiv \int d^3r{\bf A}\cdot {\bf B}$, ${\bf B}=\nabla \times {\bf A}$. But unlike the 2D case, the 3D bubbles always seem to ultimately merge, regardless of the signs of helicity and axis orientations. 

The difference between the 2D and the 3D cases must be due to the fact that in 2D we have infinitely many robust invariants $I$. In 3D, $I$ and $J$ are invariant (and even defined) only under axisymmetry. The only generic robust invariant in 3D is, apparently, the helicity. Then, if the total helicity is zero, the magnetic field has no reason to stay, and indeed it does seem to decay to zero. More work is needed here.  We can only show what we have seen:

\begin{itemize}

\item Fig.4. Just like the 2D non-circular bubbles relax into circular bubbles, in 3D, the non-axisymmetric bubbles relax into axisymmetric. 

\item Fig.5. Just like the 2D bubbles of the same polarity merge into a single circular bubble, in 3D, bubbles of the same helicity merge into a single axisymmetric bubble, conserving helicity but decreasing energy. Fig.5. shows the merger of the bubbles with the same helicity and with parallel axes\footnote{ We define the axis of the bubble to point in the direction of the magnetic field on the axis}.

\item Fig.6. The merger of the bubbles with the same helicity and with anti-parallel axes.

\item Fig.7. The 2D bubbles of the opposite polarity repel. In 3D, the opposite helicity bubbles still merge. If the total helicity is zero, apparently a total destruction of the magnetic field occurs. Fig.7 shows annihilation of the opposite helicity bubbles with anti-parallel axes. 

\end{itemize}

\section{Summary}

We have shown (tentatively, numerically) that non-singular axisymmetric equilibria are weakly dissipative attractors. The unsolved problems are:

1. Do other weakly dissipative (that is non-singular) attractors exist?

2. Do the weakly dissipative attractors form (and thus exist) in nature?

\begin{figure}
\plottwo{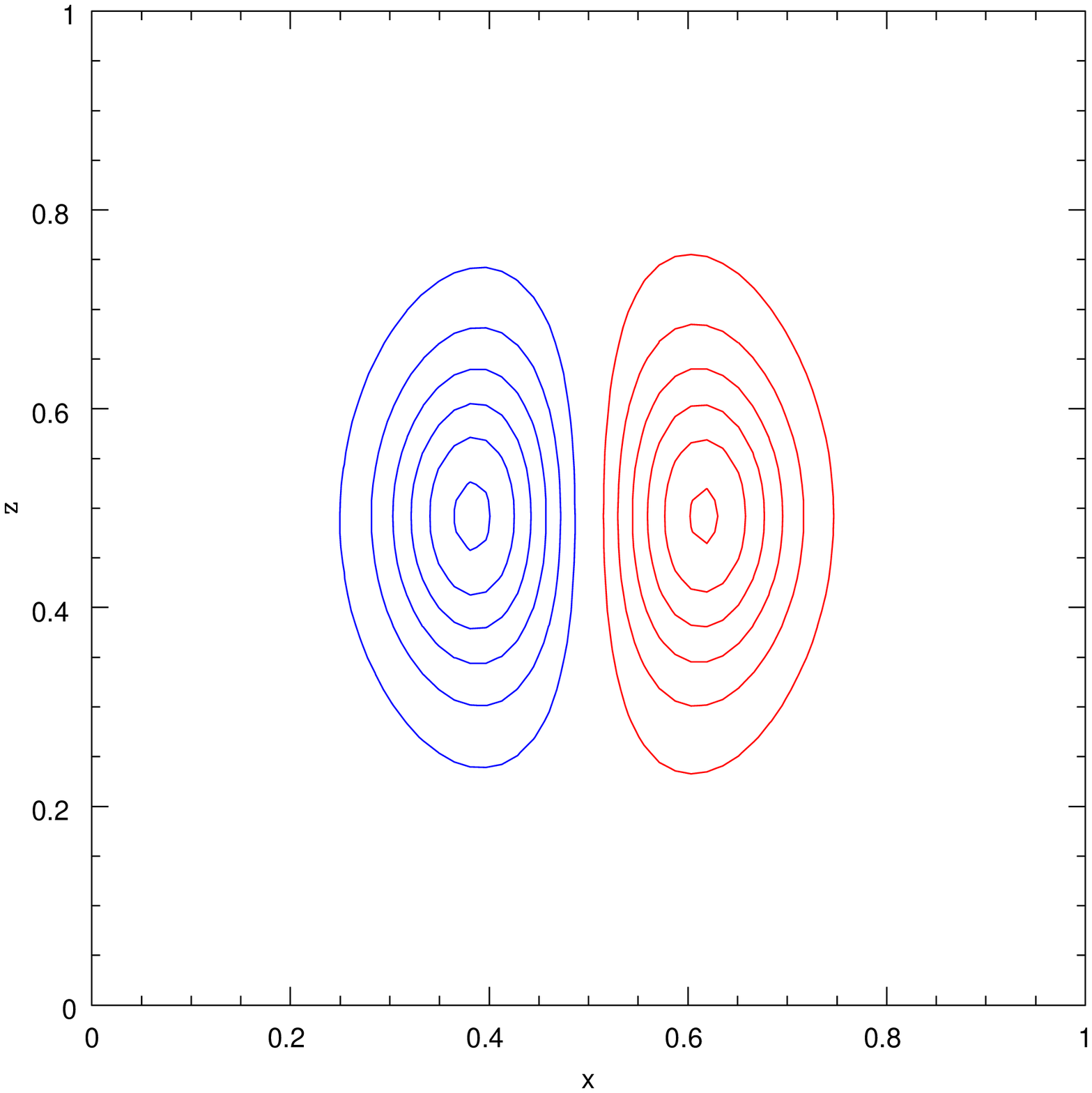}{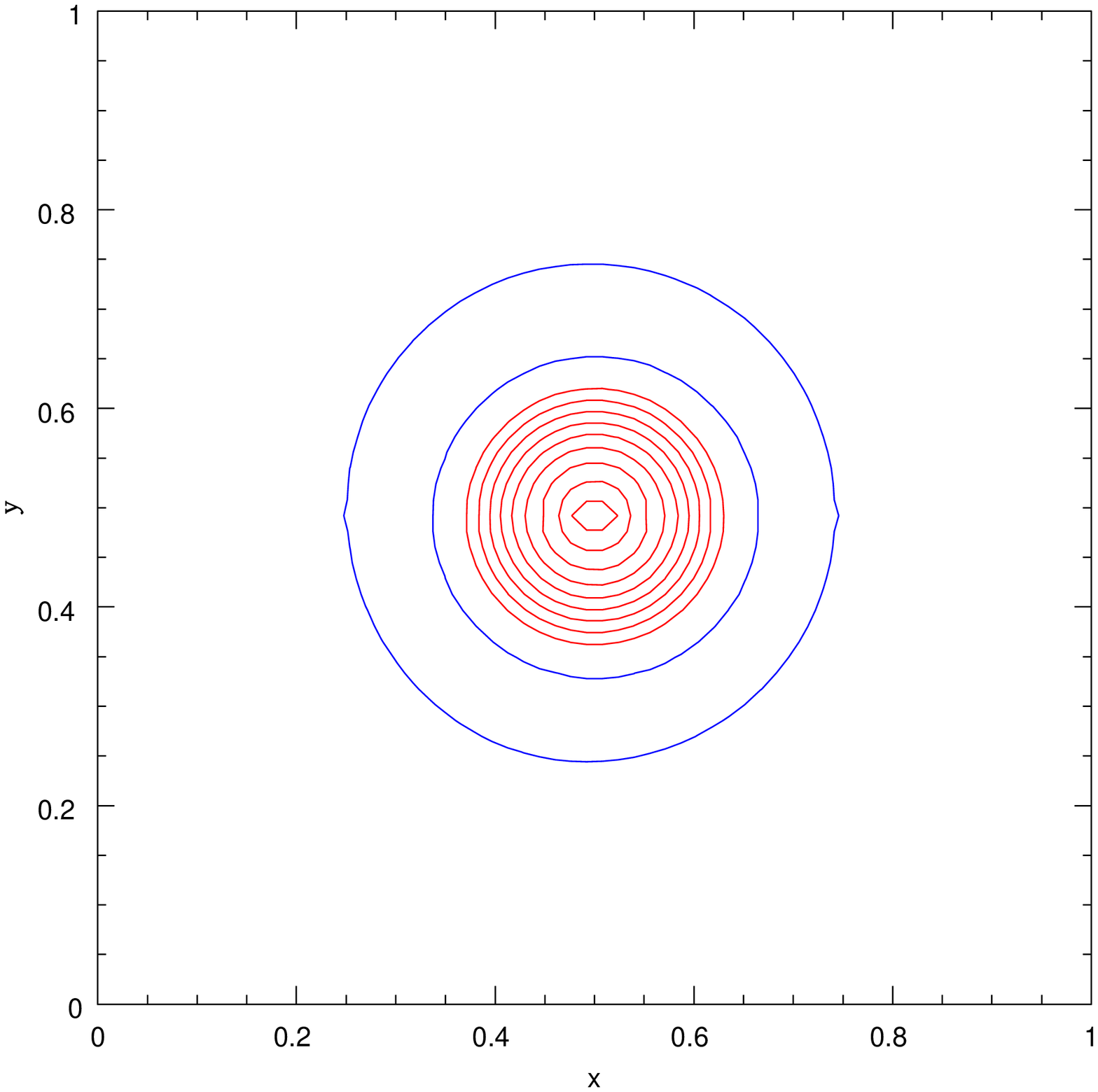}
\plottwo{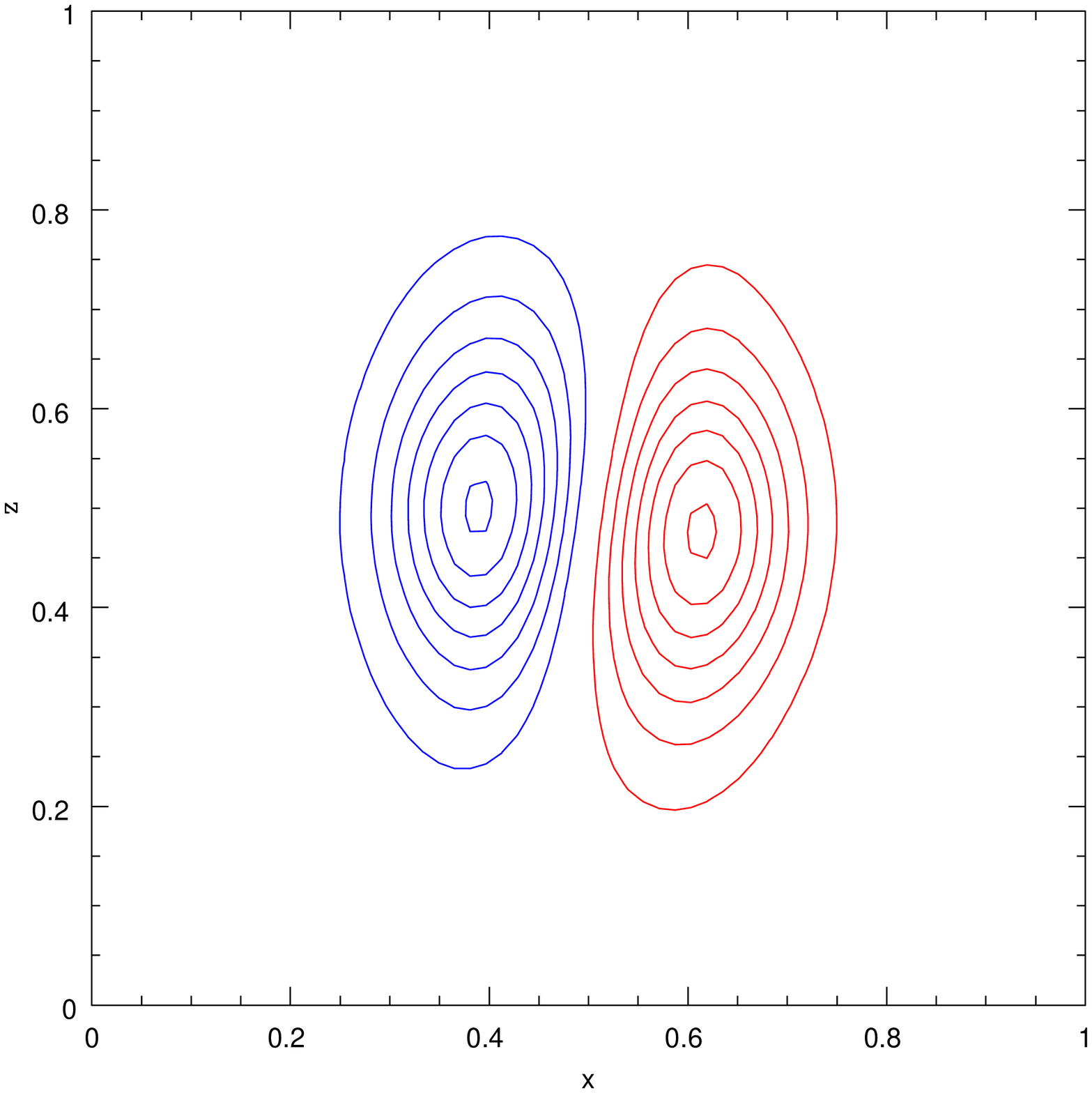}{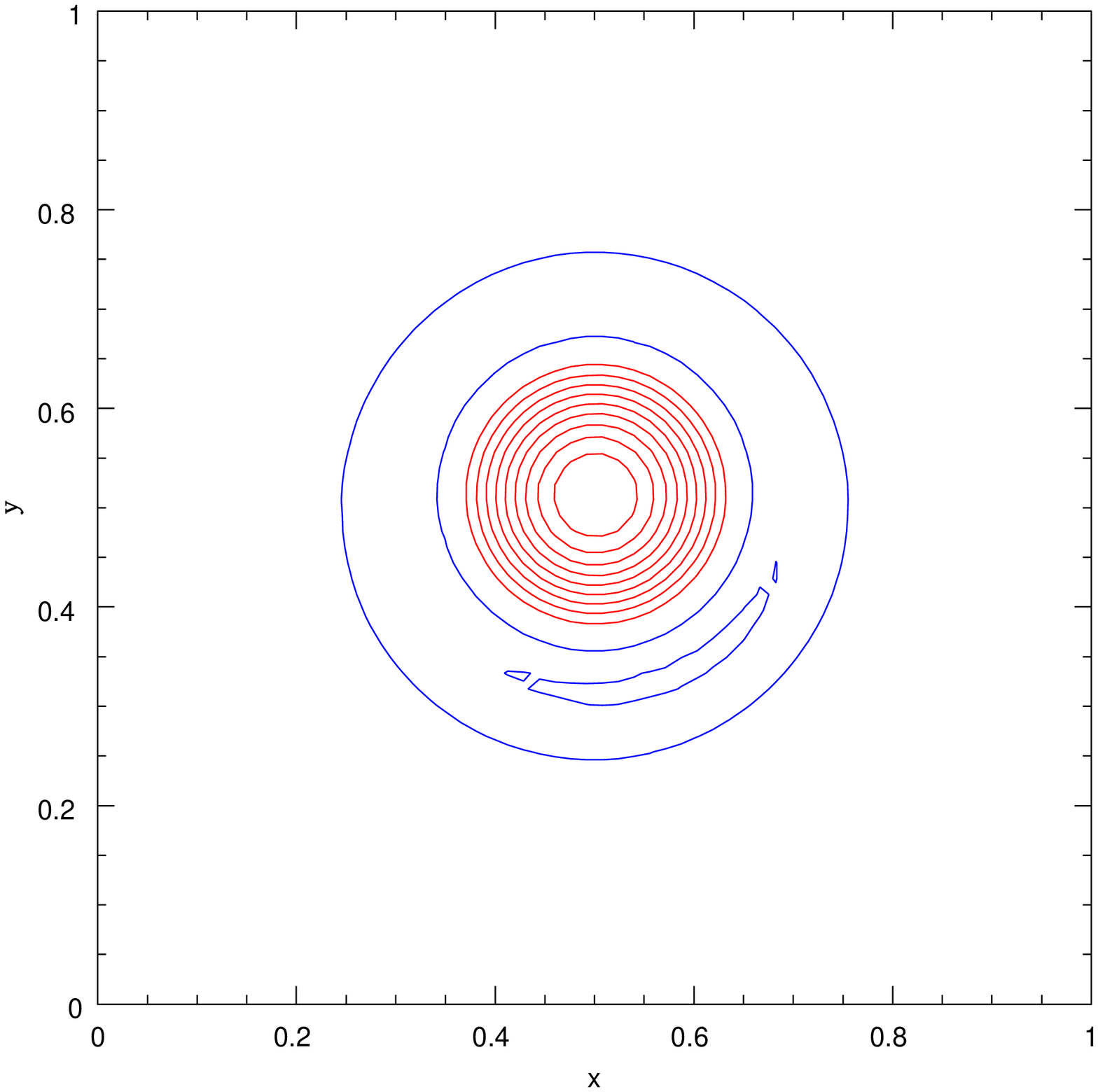}

\caption{{\it Upper Left}: Initial isolines of $B_y$ in the plane $y=0.5$. Different colors indicate different signs of $B_y$ {\it Upper Right}: Initial $B_z$ in the plane $z=0.5$. {\it Lower Left}: Final $B_y$ {\it Lower Right}: Final $B_z$ }
\end{figure}

\begin{figure}
\plottwo{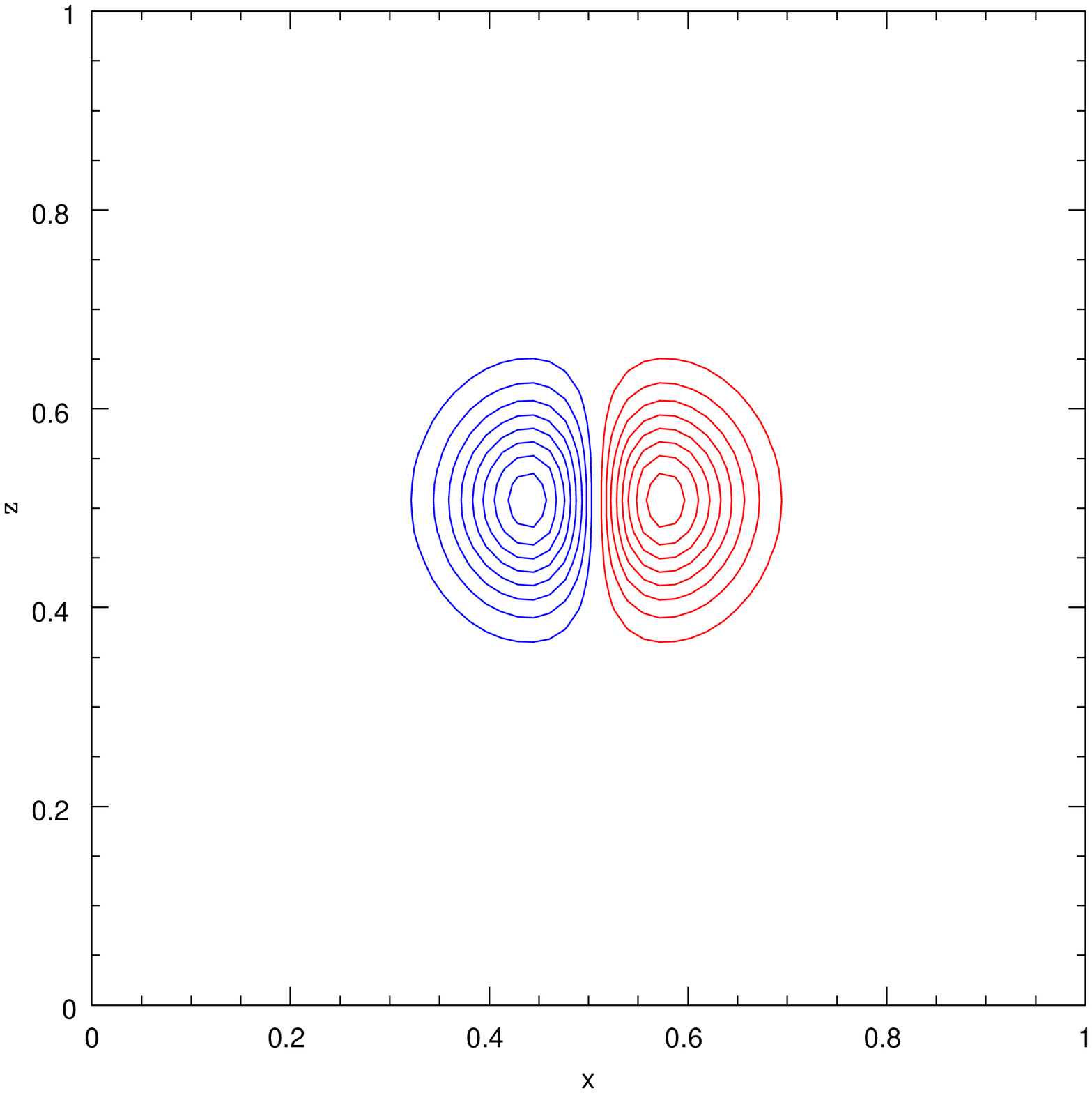}{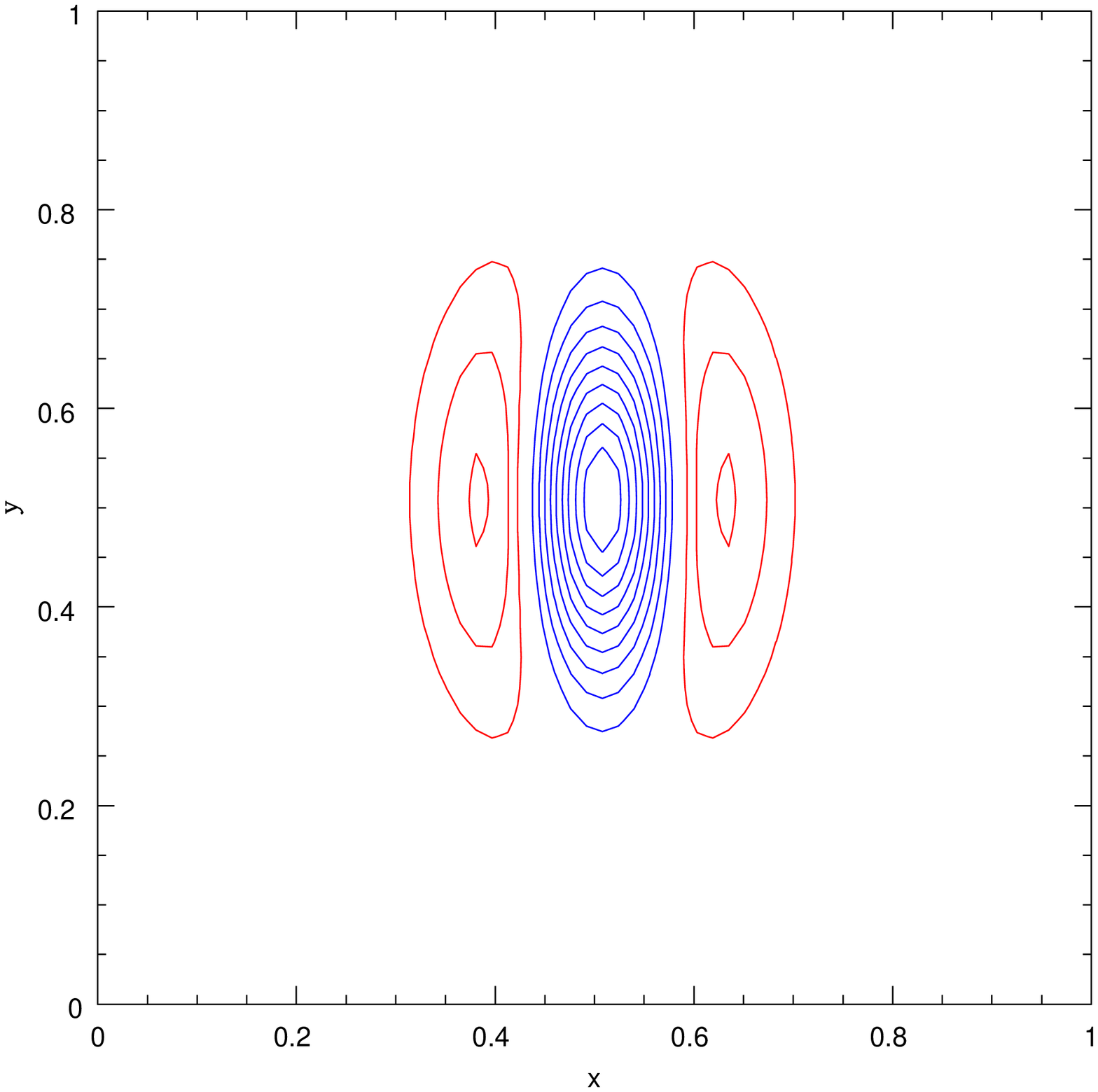}
\plottwo{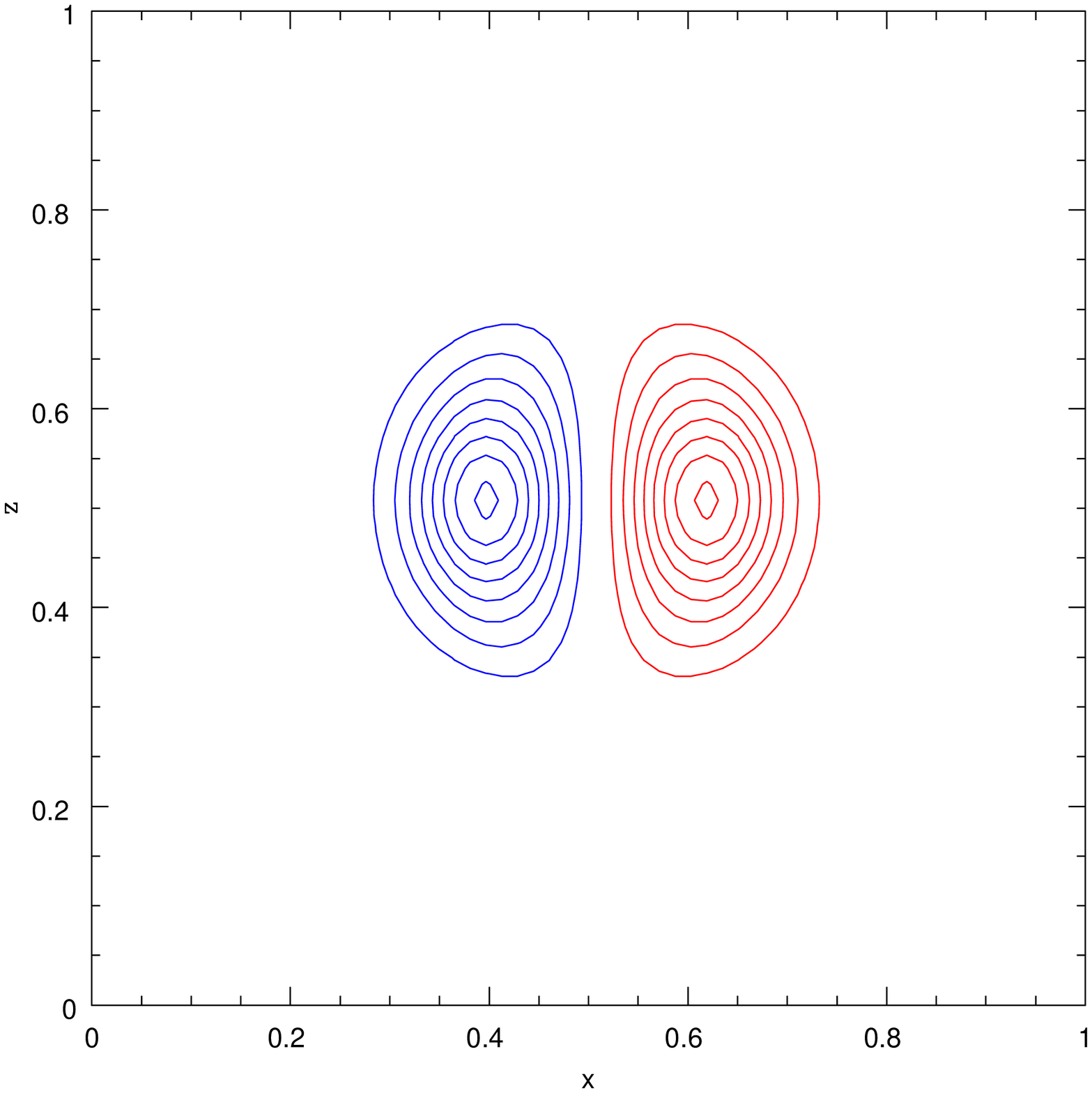}{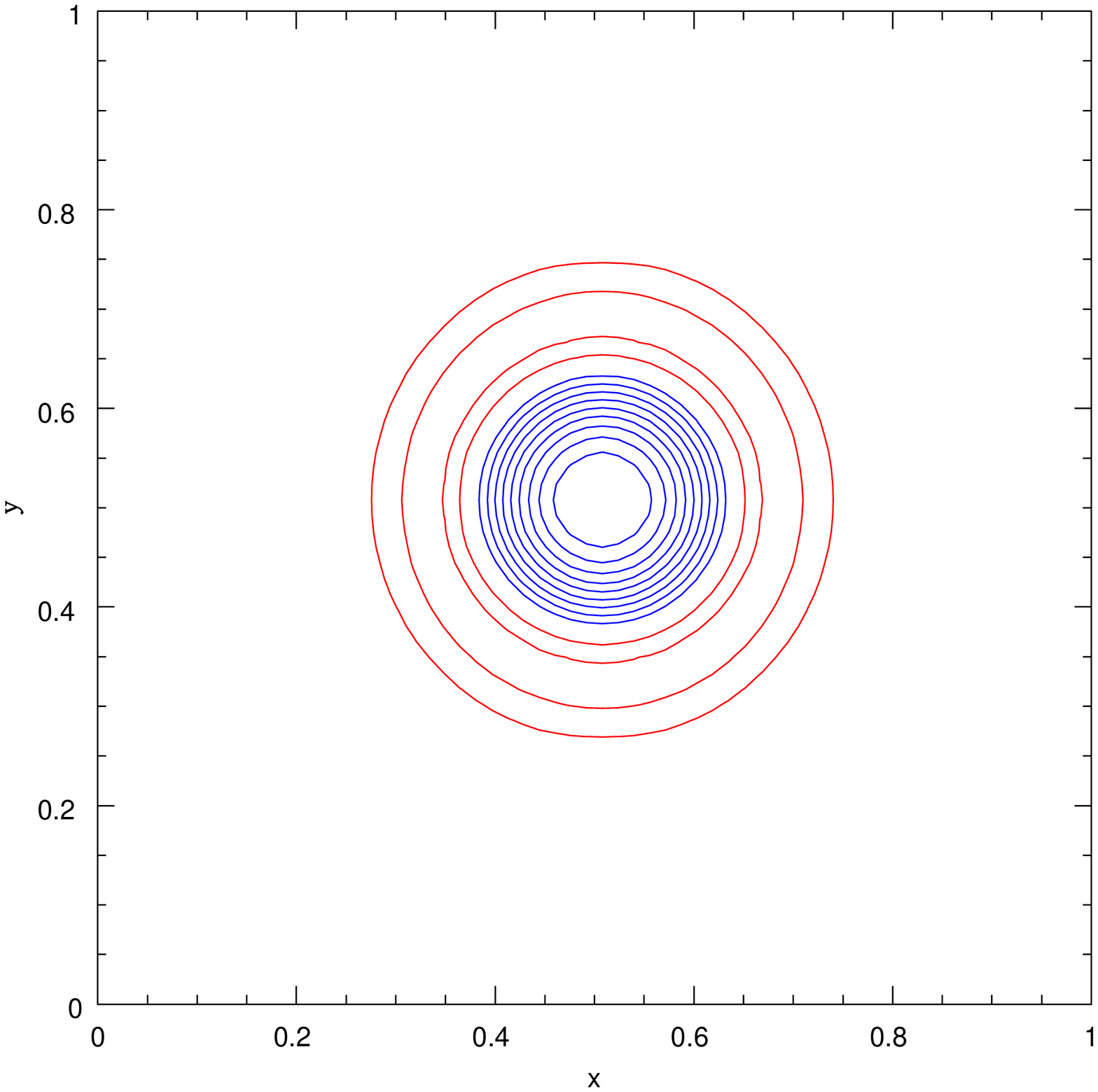}

\caption{Same as fig.3. The energy decreases by 15\% .}
\end{figure}

\begin{figure}
\plottwo{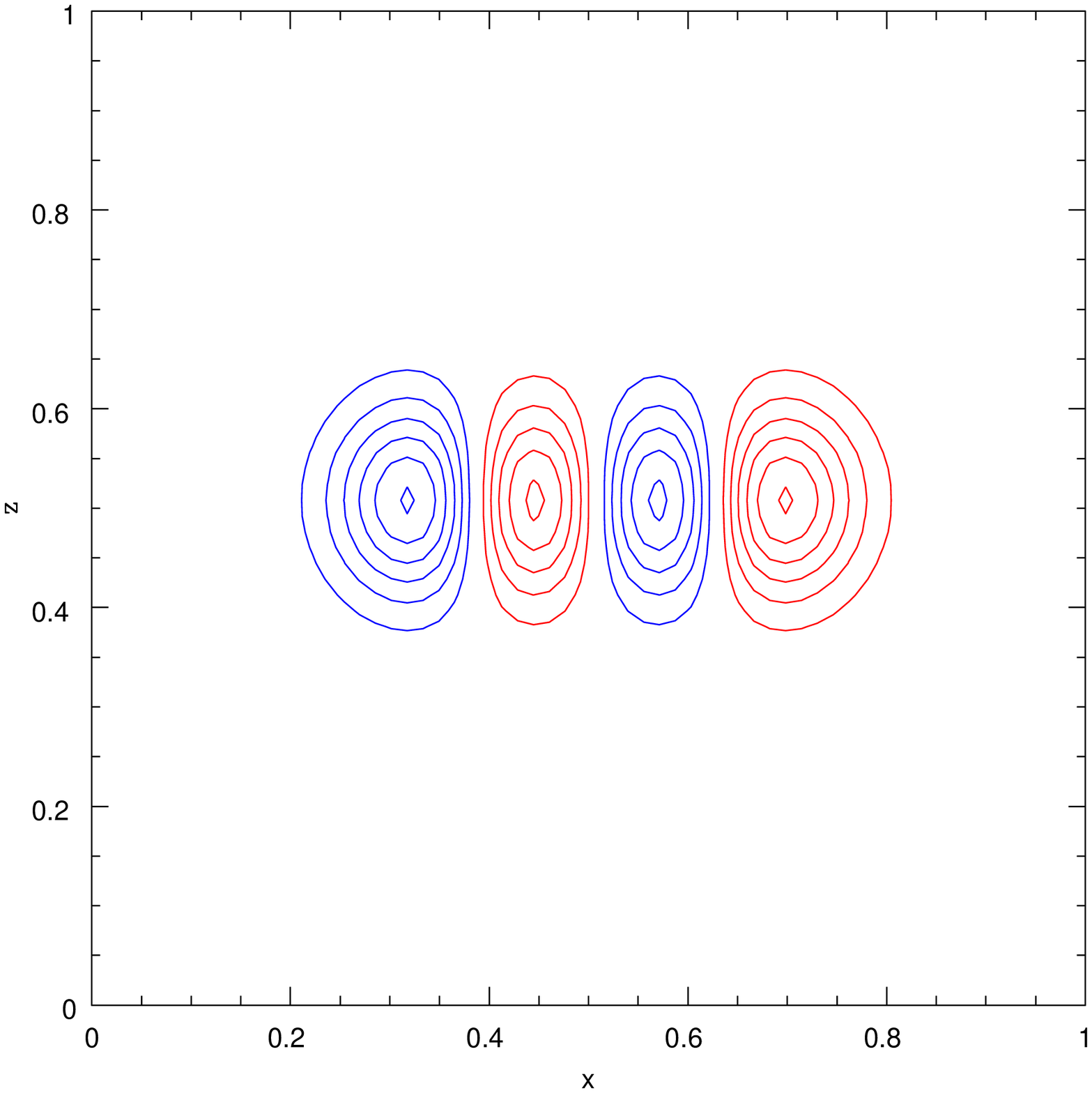}{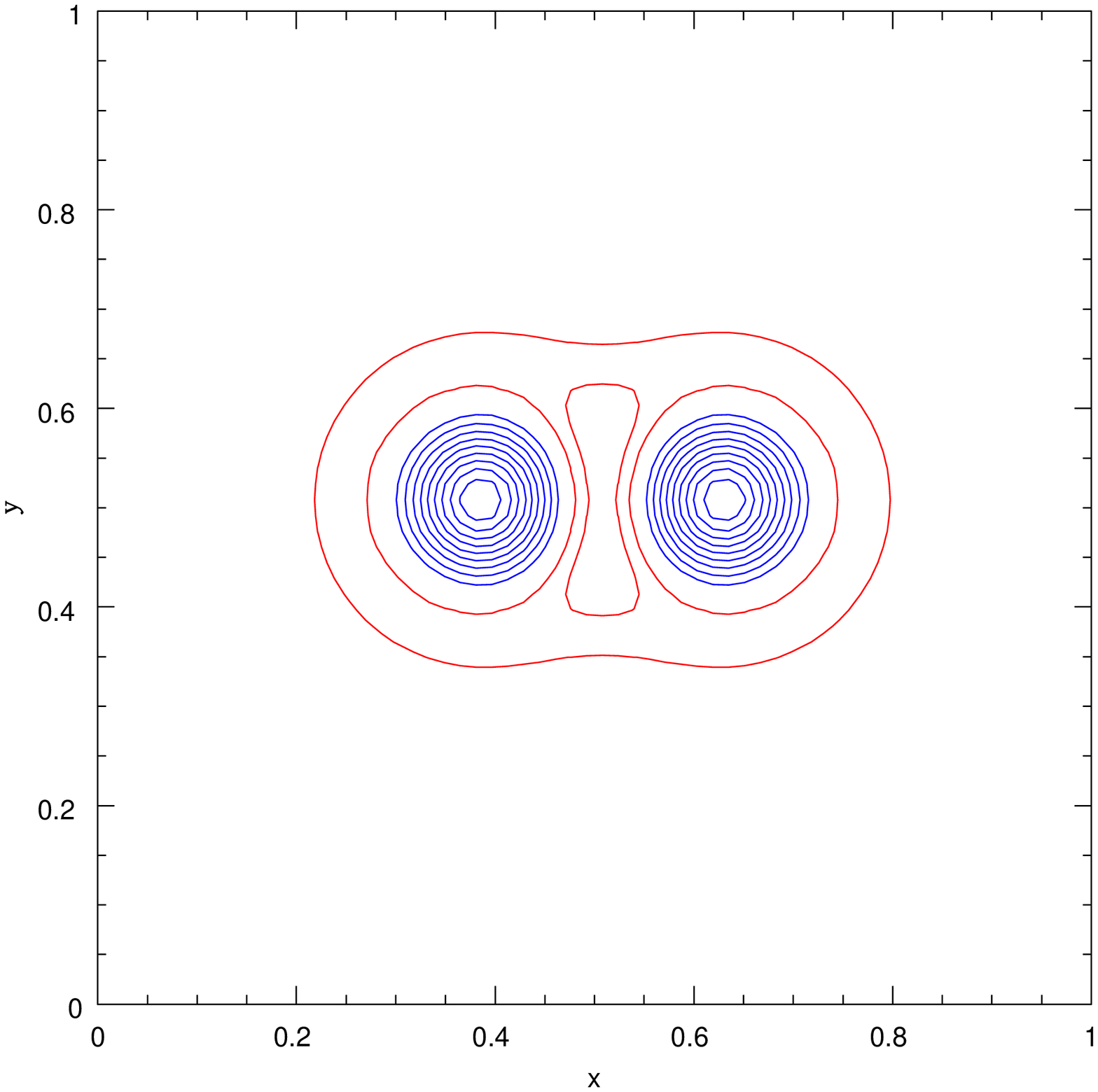}
\plottwo{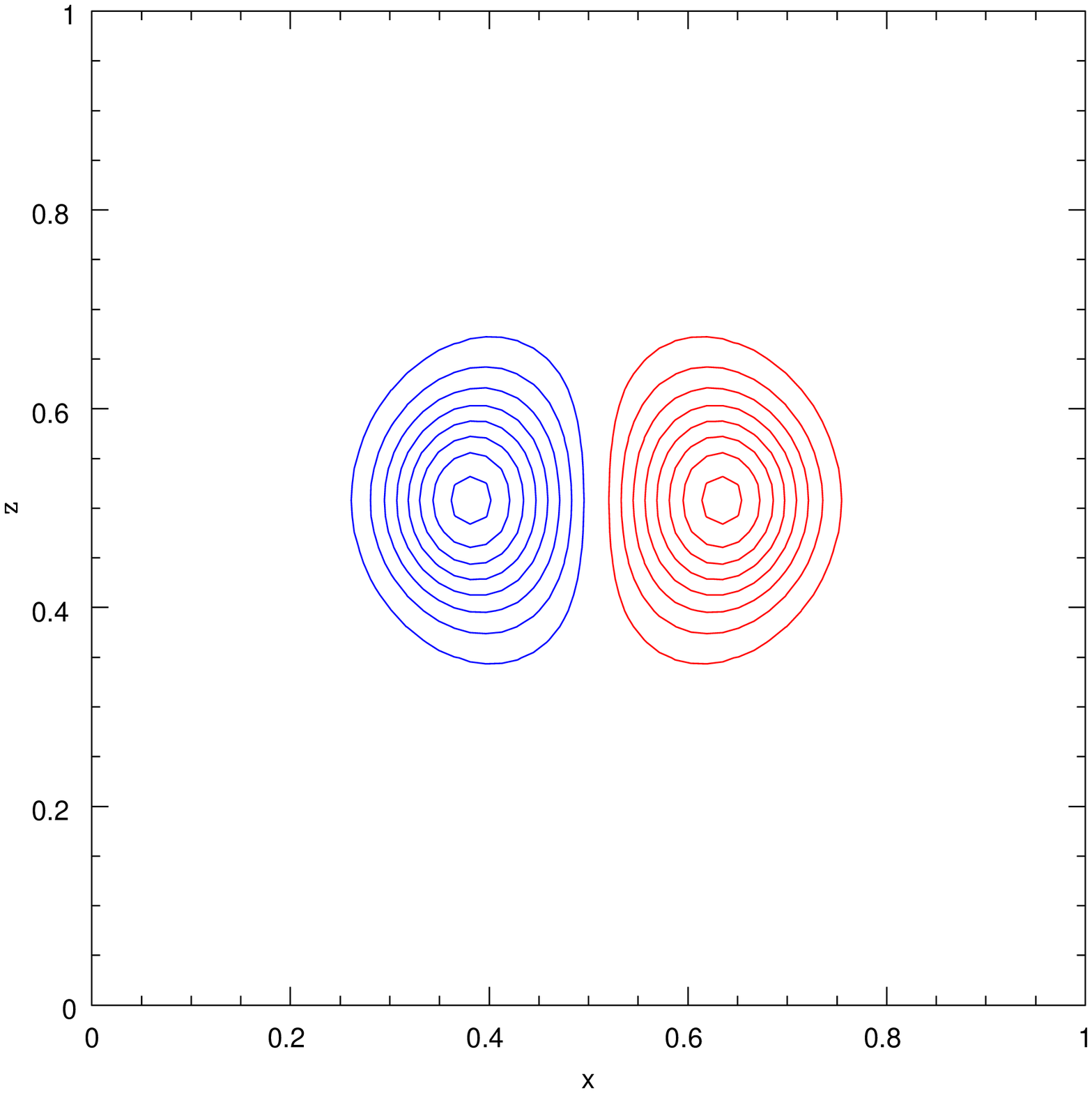}{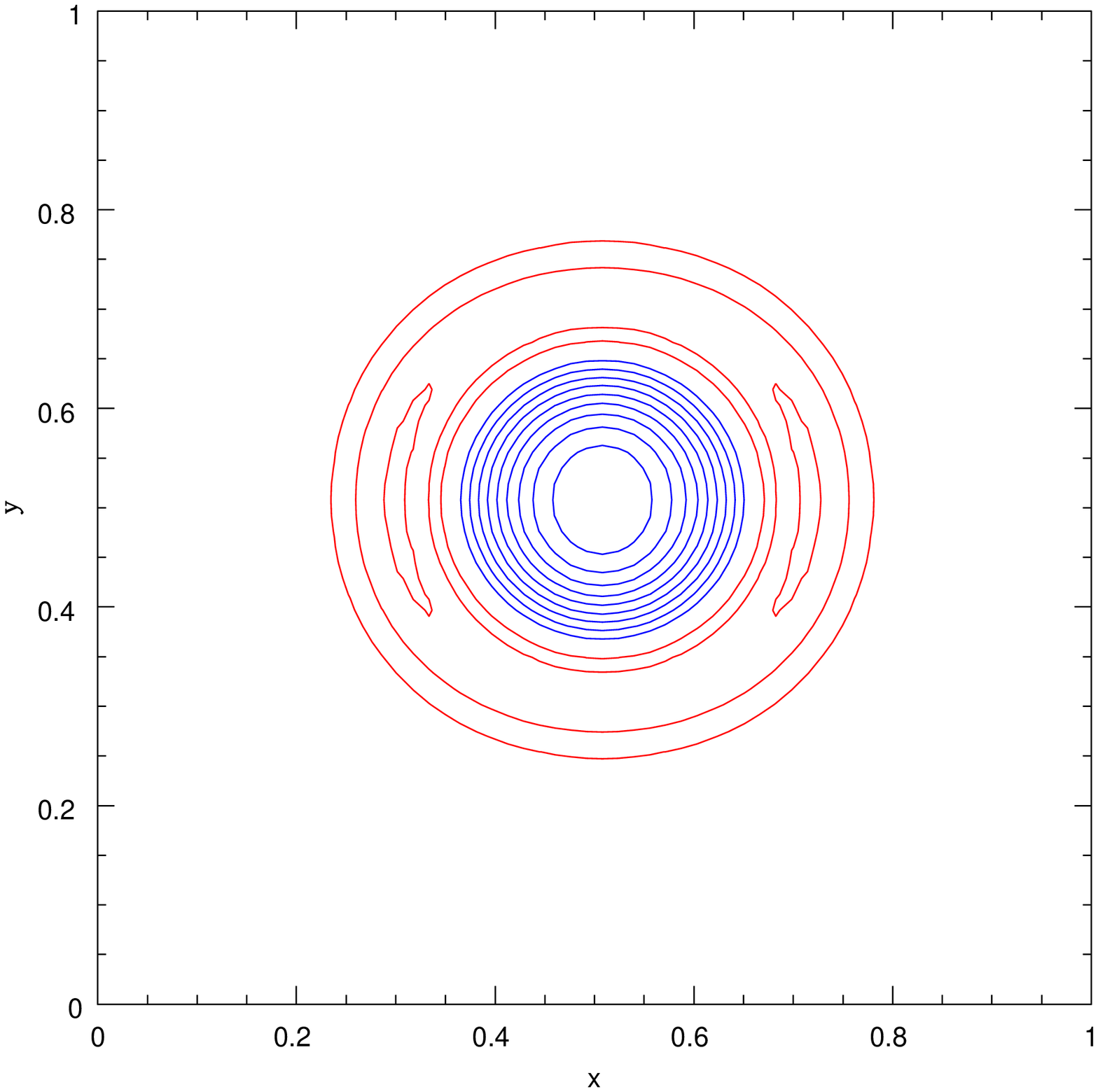}

\caption{Same as fig.3. The energy decreases by 34\% .}
\end{figure}

\begin{figure}
\plottwo{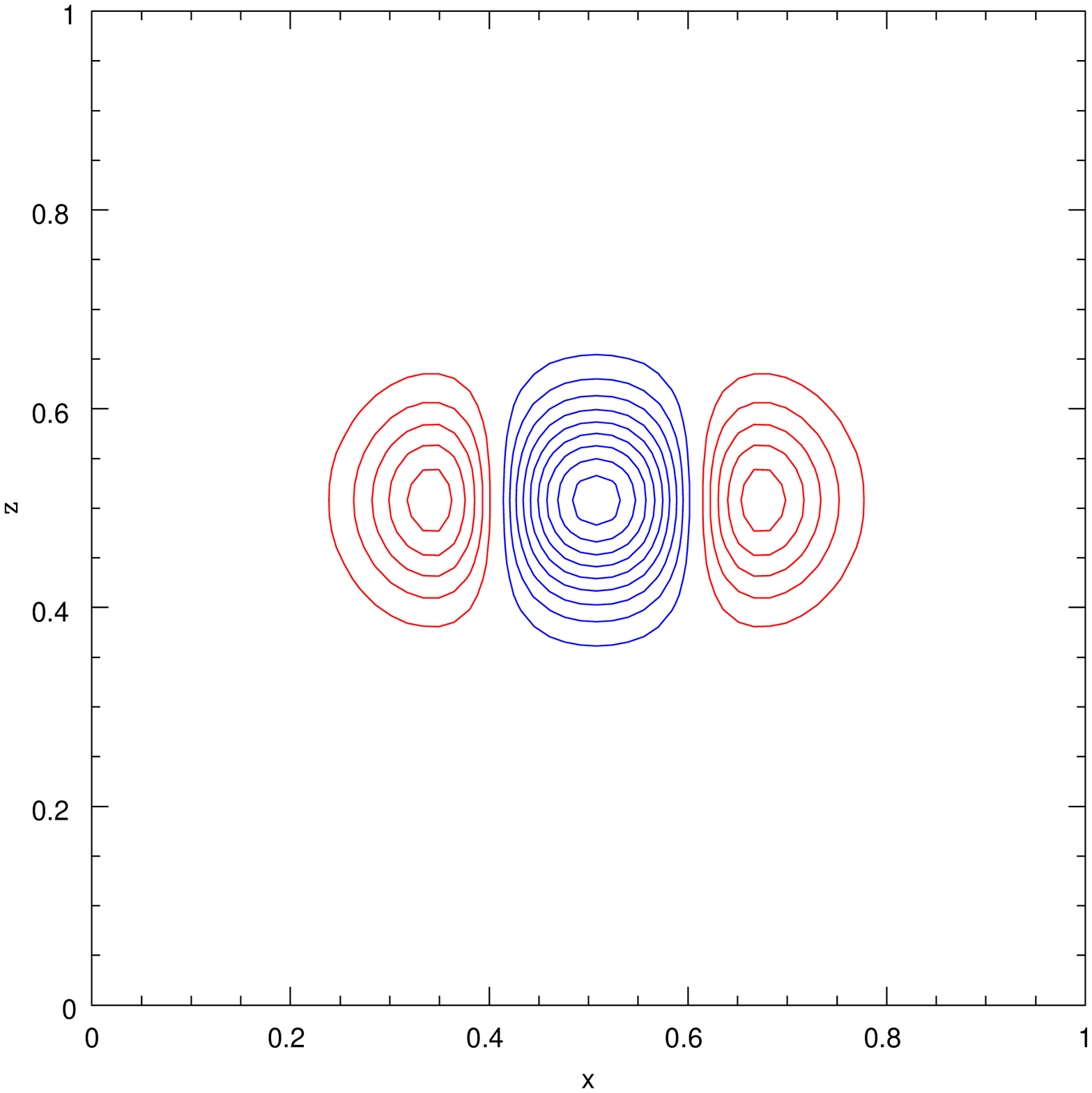}{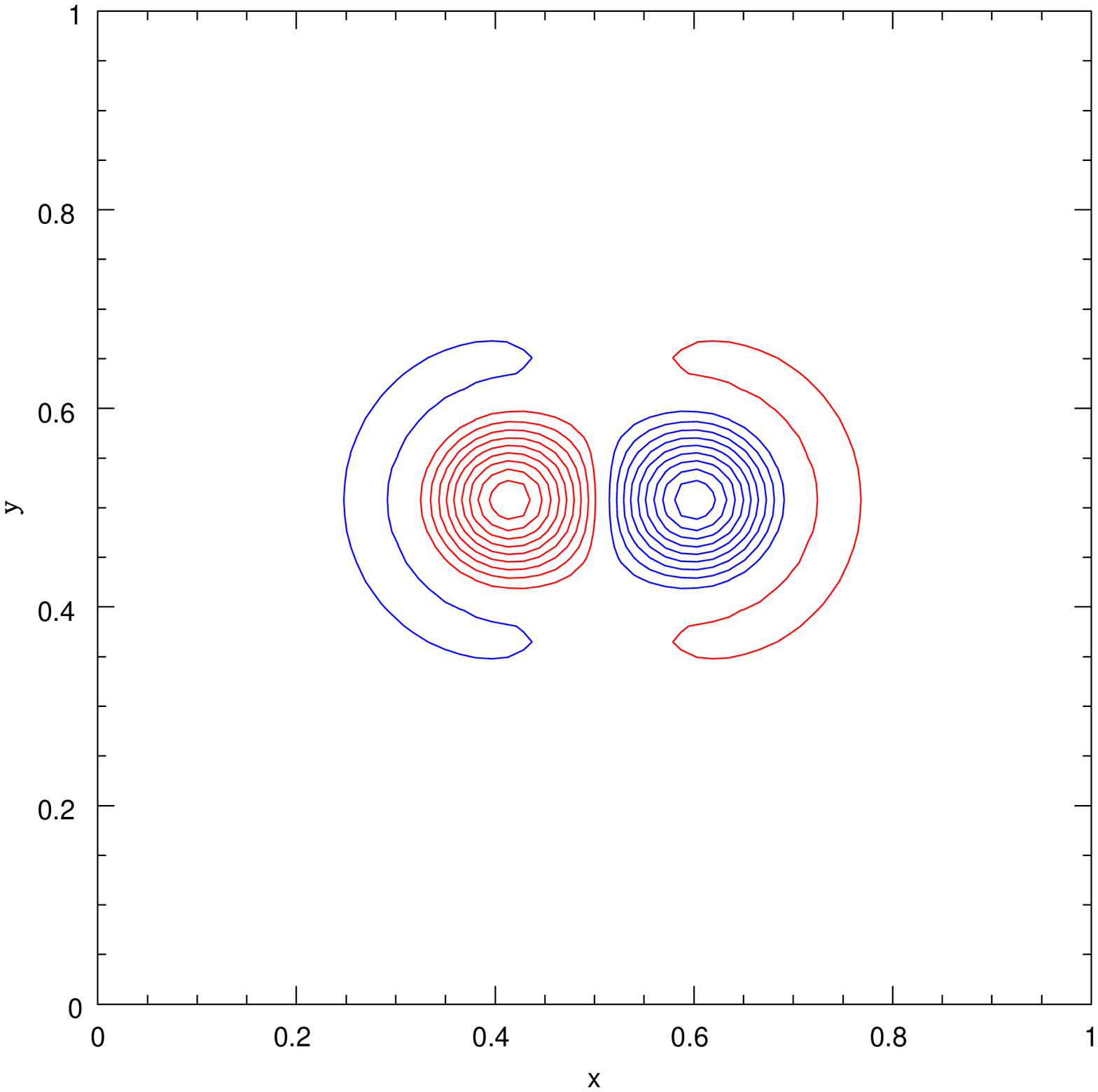}
\plottwo{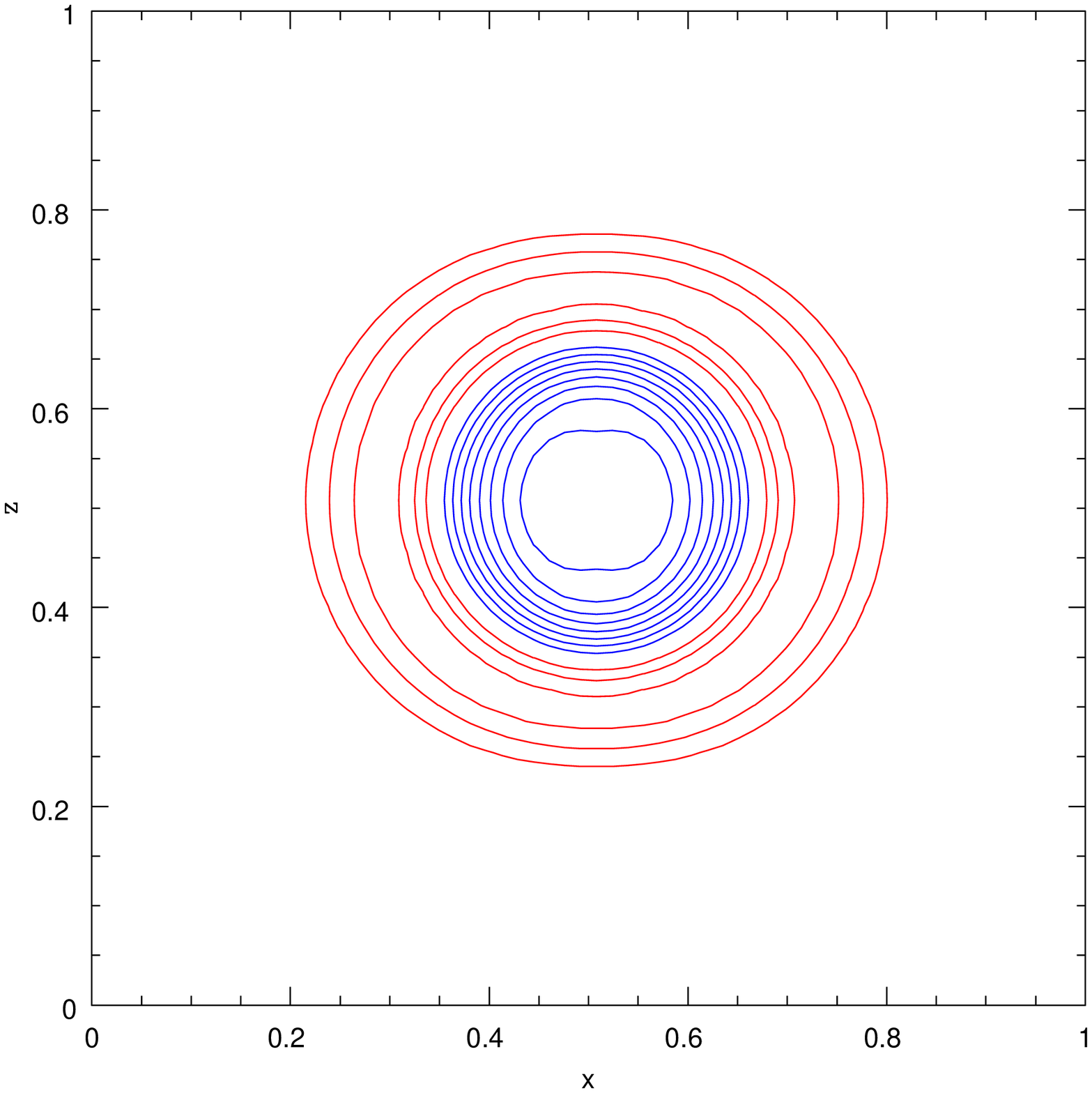}{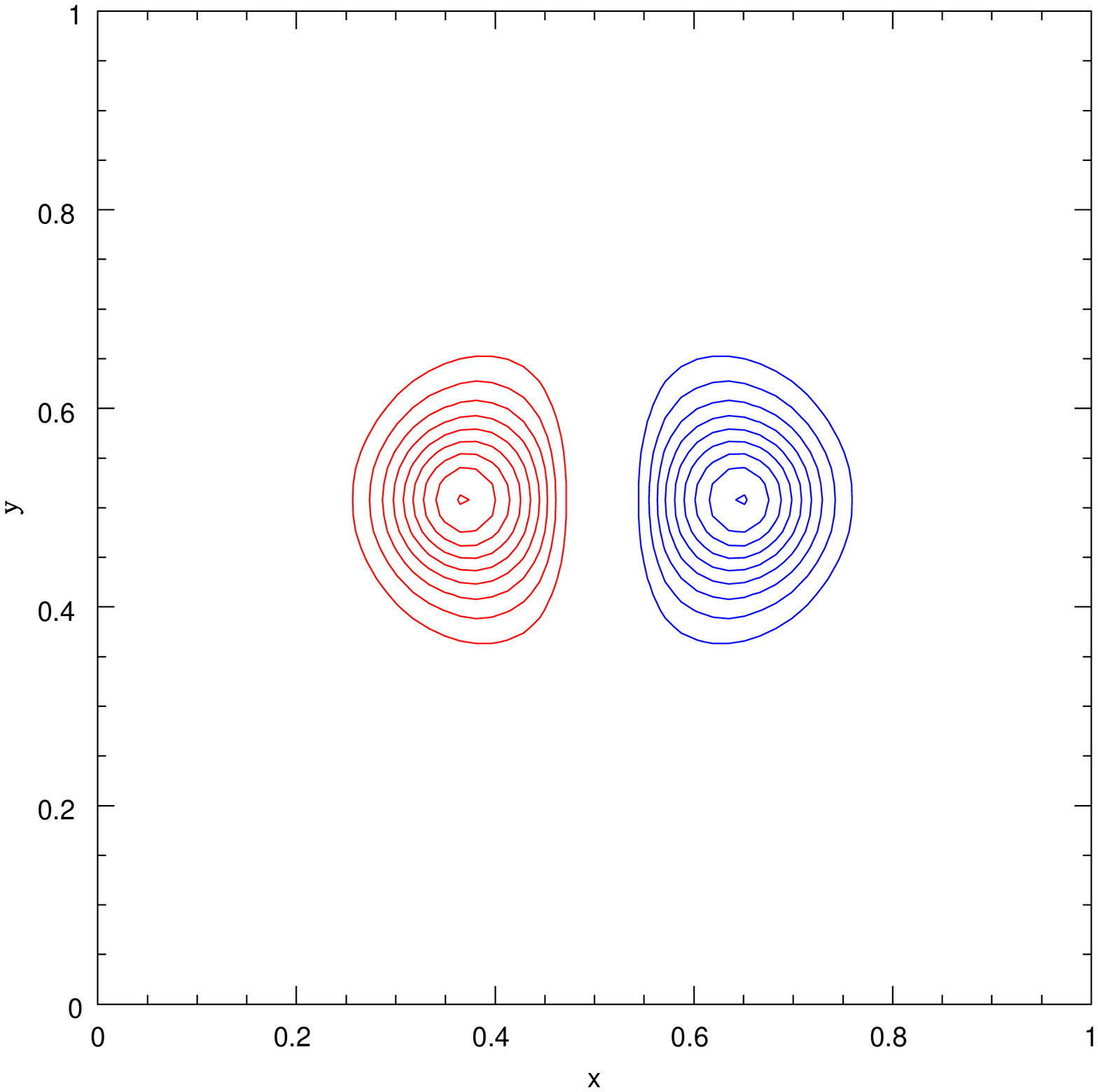}

\caption{Same as fig.3. The energy decreases by 35\% .}
\end{figure}

\begin{figure}
\plottwo{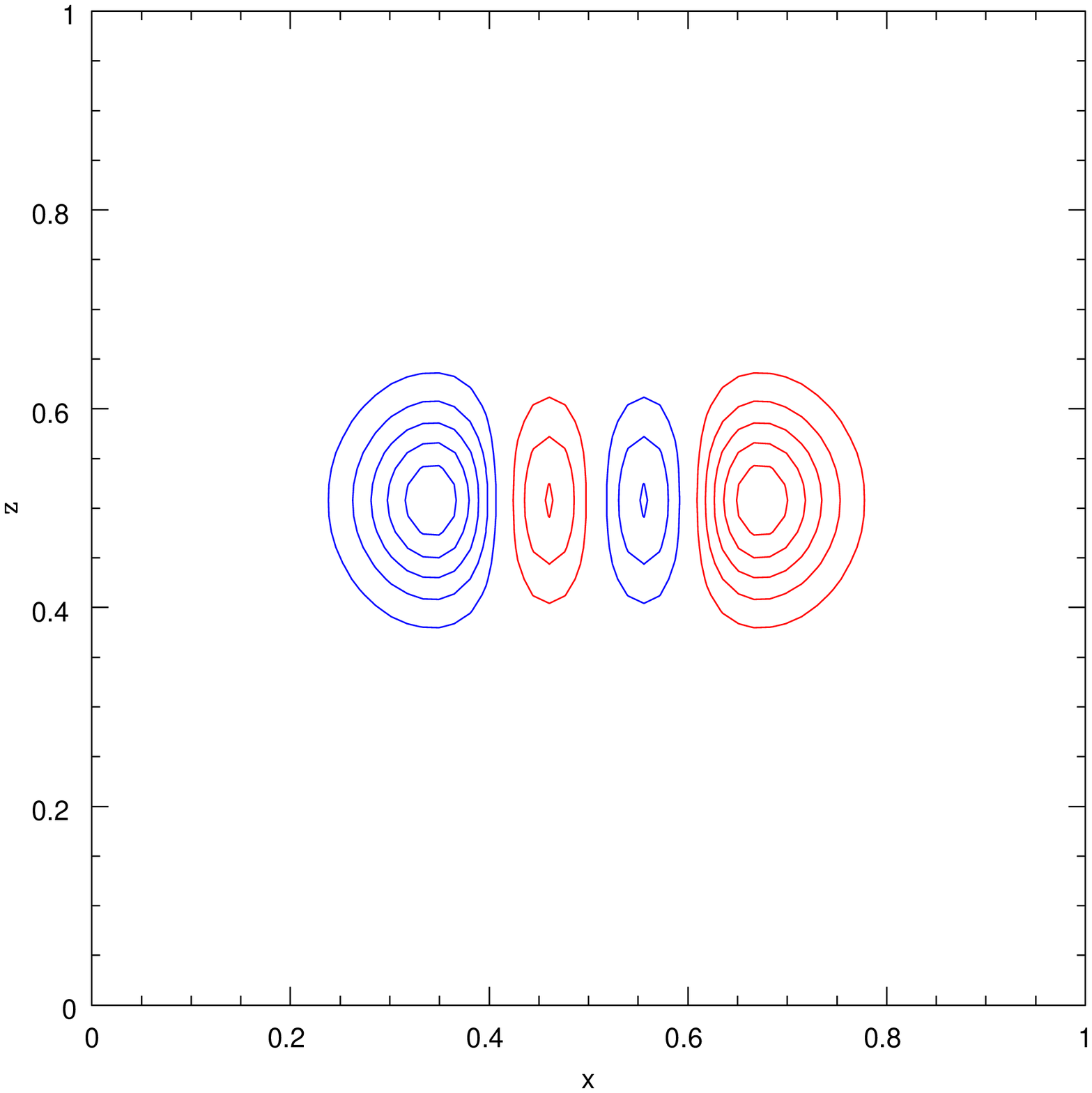}{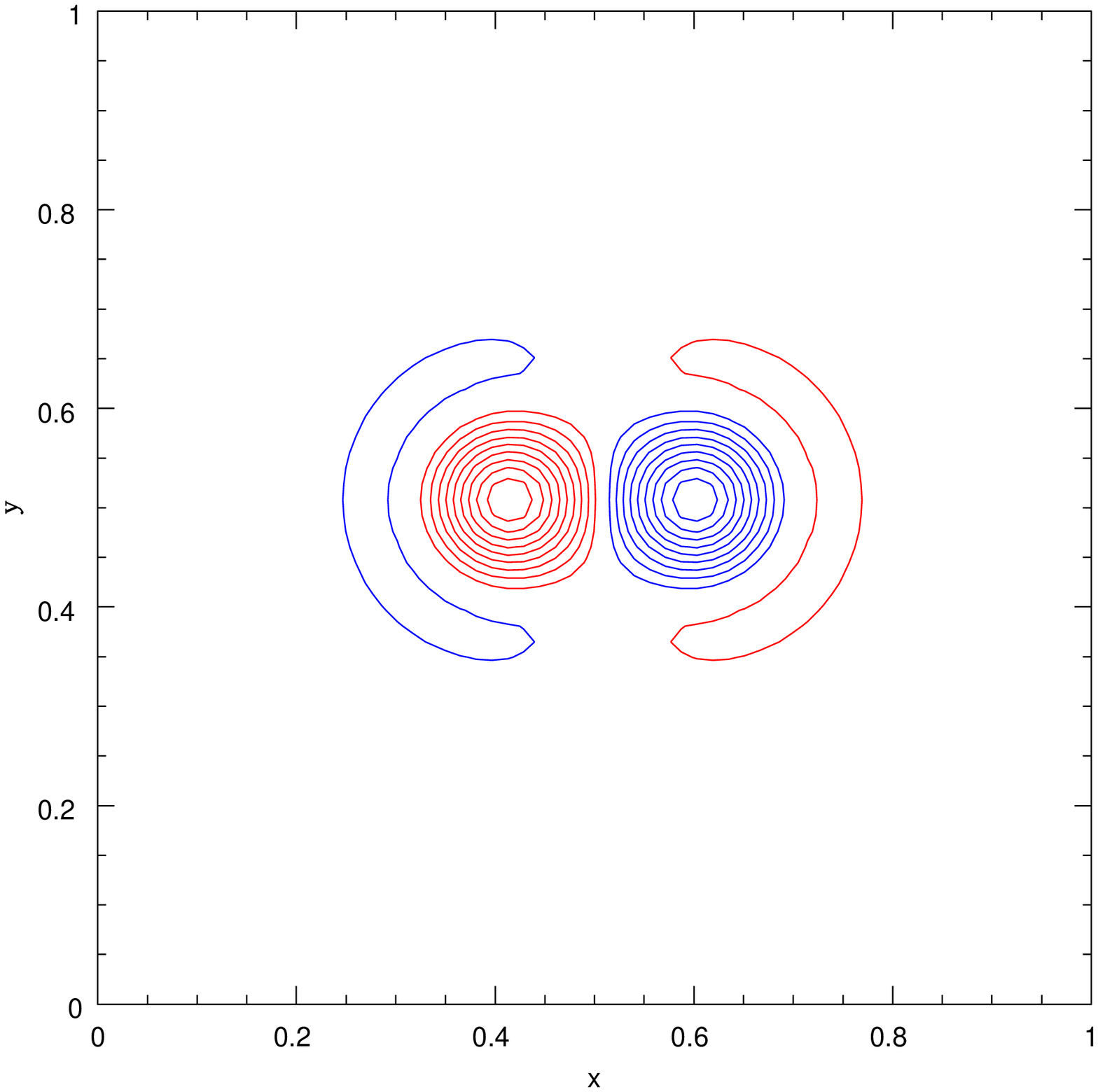}
\plottwo{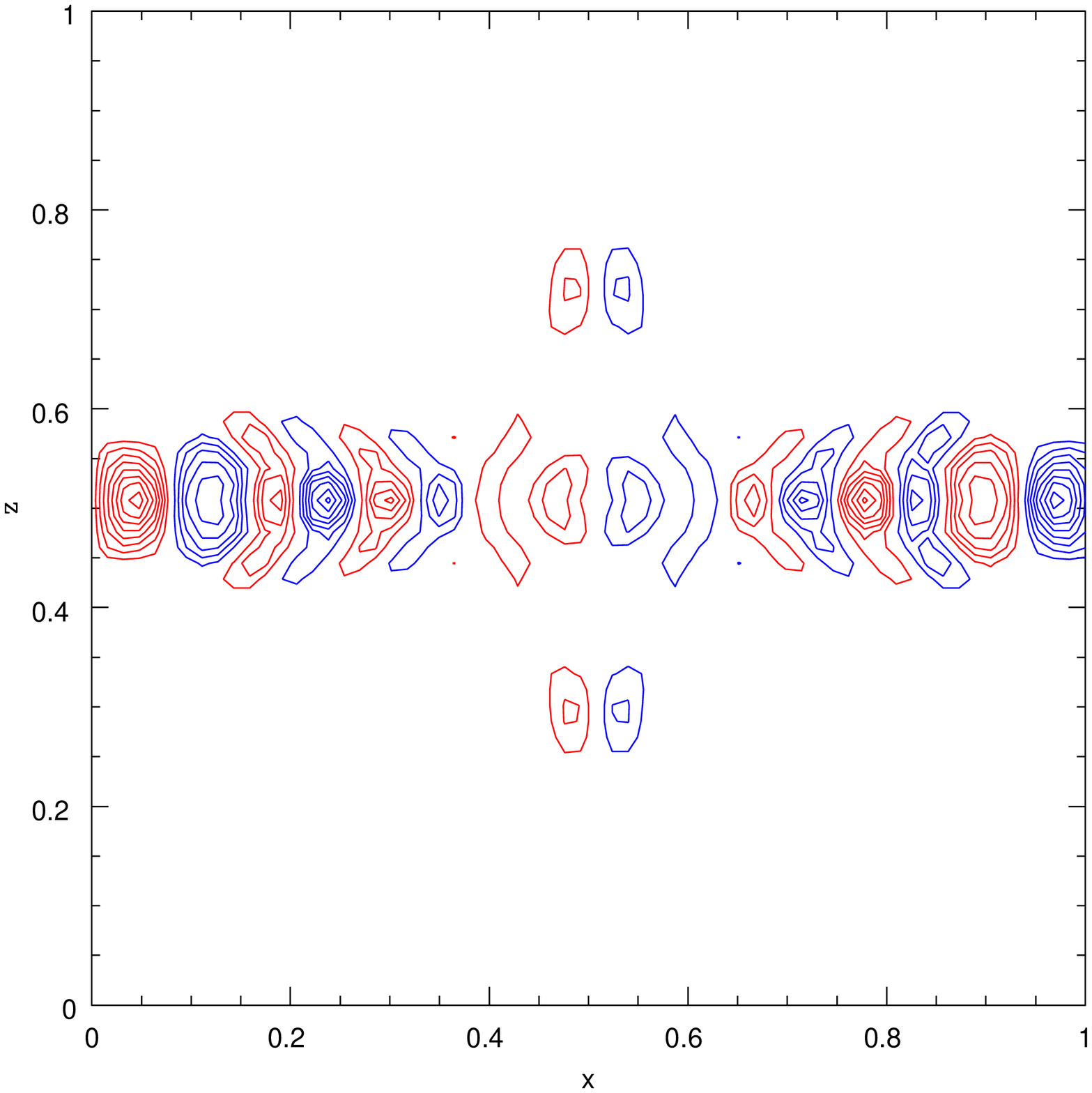}{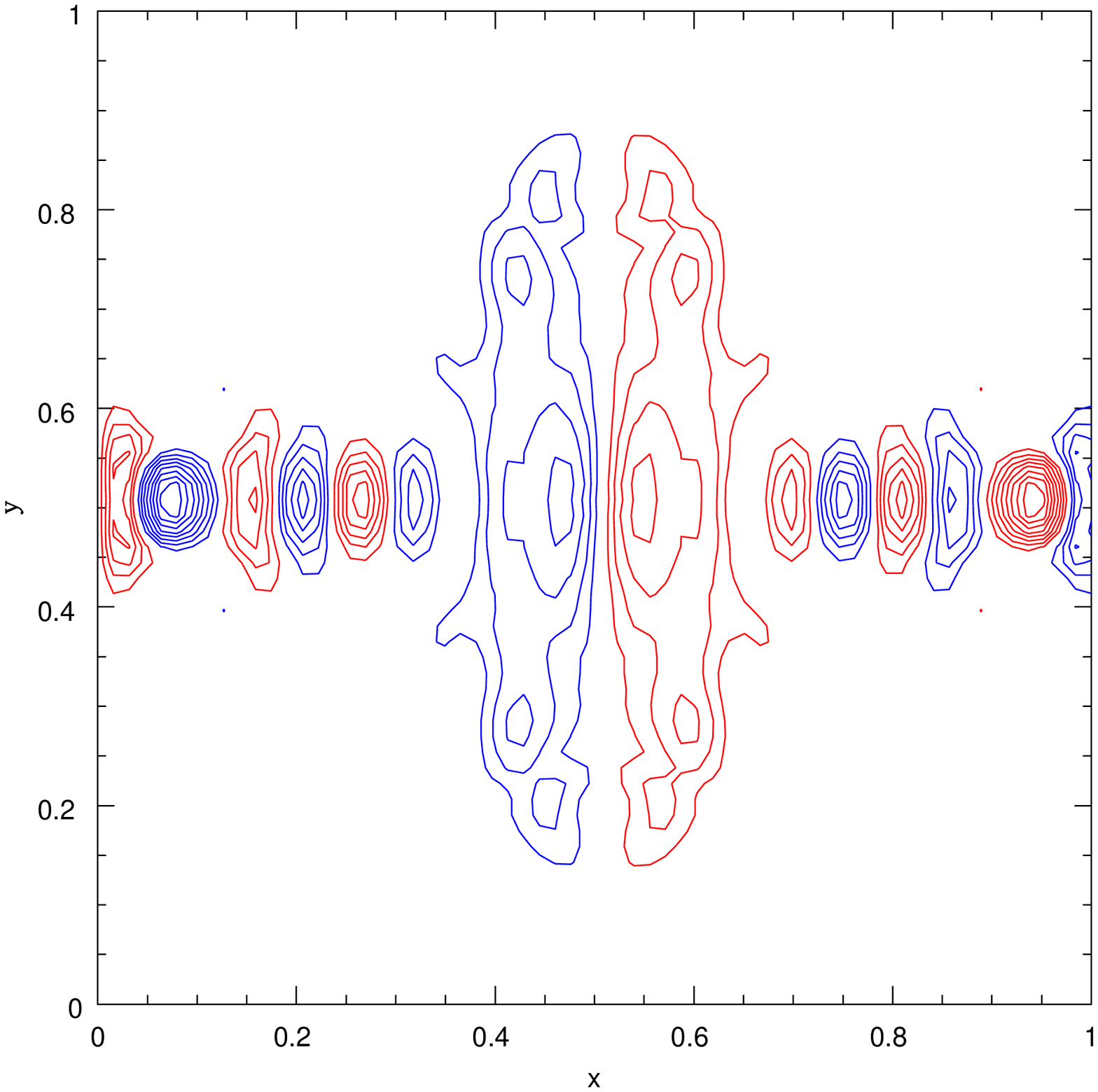}

\caption{Same as fig.3. The energy decreases by factor 13.}
\end{figure}

\end{document}